\documentclass[aps,prb,showpacs,twocolumn,citeautoscript,superscriptaddress,footinbib,bibnotes]{revtex4-1}
\usepackage[dvips]{graphicx}
\usepackage{dcolumn}
\usepackage{bm}
\usepackage{fancyhdr}
\usepackage{placeins}
\usepackage{ifthen}
\usepackage{eurosym}
\usepackage{float}
\usepackage{longtable}
\usepackage{hyperref}
\usepackage{footnote}
\usepackage[usenames,dvipsnames]{color}

\begin{document}
\preprint{YBCF}
\title{Incommensurate magnetic structure, Fe/Cu chemical disorder and magnetic interactions in the high-temperature multiferroic YBaCuFeO$_5$}
\vspace{-7mm}
\author{M.~Morin}
\affiliation{Laboratory for Developments and Methods, Paul Scherrer Institut, 5232 Villigen PSI, Switzerland}
\author{A.~Scaramucci}
\affiliation{Laboratory for Developments and Methods, Paul Scherrer Institut, 5232 Villigen PSI, Switzerland}
\affiliation{Materials Theory, ETH Z\"{u}rich, 8093 Z\"{u}rich, Switzerland}
\author{M.~Bartkowiak}
\author{E.~Pomjakushina}
\affiliation{Laboratory for Developments and Methods, Paul Scherrer Institut, 5232 Villigen PSI, Switzerland}
\author{G.~Deng}
\affiliation{Laboratory for Developments and Methods, Paul Scherrer Institut, 5232 Villigen PSI, Switzerland}
\affiliation{Bragg Institute, Australia Nuclear Science and Technology Organization, New Illawarra Road, Lucas Height, New South Wales 2233, Australia}
\author{D.~Sheptyakov}
\author{L.~Keller}
\affiliation{Laboratory for Neutron Scattering and Imaging, Paul Scherrer Institut, 5232 Villigen PSI, Switzerland}
\author{J.~Rodriguez-Carvajal}
\affiliation{Institut Laue Langevin, BP 156, 6, rue Jules Horowitz, 38042 Grenoble Cedex 9, France}
\author{N.A.~Spaldin}
\affiliation{Materials Theory, ETH Z\"{u}rich, 8093 Z\"{u}rich, Switzerland}
\author{M.~Kenzelmann}
\author{K.~Conder}
\author{M.~Medarde}
\email{marisa.medarde@psi.ch}
\affiliation{Laboratory for Developments and Methods, Paul Scherrer Institut, 5232 Villigen PSI, Switzerland}

\date{\today}

\begin{abstract}
Motivated by the recent observations of incommensurate magnetic order and electric polarization in YBaCuFeO$_5$ up to temperatures $T_{N2}$ as high as 230K \cite{Kundys_09,Kawamura_10} we report here for the first time a model for the incommensurate magnetic structure of this material that we complement with ab-initio calculations of the magnetic exchange parameters. Using neutron powder diffraction we show that the appearance of polarization below $T_{N2}$ is accompanied by the replacement of the high temperature collinear magnetic order by a circular inclined spiral with propagation vector \textbf{k$_i$}=(1/2, 1/2, 1/2$\pm$\textit{q}). Moreover, we find that the polarization approximately scales with the modulus of the magnetic modulation vector \textit{q} down to the lowest temperature investigated (3 $\sim$K). Further, we observe occupational Fe/Cu disorder in the FeO$_5$-CuO$_5$ bipyramids, although a preferential occupation of such units by Fe-Cu pairs is supported by the observed magnetic order and by density functional calculations. We calculate exchange coupling constants for different Fe/Cu distributions and show that, for those containing Fe-Cu dimers, the resulting magnetic order is compatible with the experimentally observed collinear magnetic structure (\textbf{k$_c$}=(1/2, 1/2, 1/2), $T_{N2}$ $>$ T $>$ $T_{N1}$ = 440K). Based on these results we discuss possible origins for the incommensurate modulation and its coupling with ferroelectricity.
\end{abstract}

\maketitle

\section{Introduction}

The discovery of materials with strongly coupled magnetic and ferroelectric orders has raised a great deal of interest in view of their possible use in magnetoelectic device applications. In many such materials the spontaneous appearance of electric polarization (\textit{P}) is linked to the onset of incommensurate (ICM) magnetic order~\cite{Goto2004, Katsura2005, Cheong2007, Mostovoy2006, Dagotto2006}. This is often the signature of competing magnetic interactions ~\cite{Mochizuki2009} and hence is characterized by low ordering temperatures. As a result, their promising technological multifunctionalities such as the control of magnetism by applied electric fields usually occur at temperatures too low for most practical applications.

To date there are only two known examples of switchable, magnetism-driven ferroelectricity at zero field and temperatures above 200K. One is cupric oxide CuO \cite{Kimura_08,Giovanetti_11}, where the low monoclinic symmetry promotes frustrated magnetic interactions. A spiral magnetic multiferroic phase results in the limited temperature range of 213 to 230K \cite{Kimura_08}. The second example, which has received much less attention due to its unavailability as a single crystal or thin film, is the layered perovskite YBaCuFeO$_{5}$. This  material was first synthesized in 1988 \cite{Rakho_88}, one year after the discovery of high-temperature superconductivity in YBa$_2$Cu$_3$O$_{6+x}$ \cite{Chu_87}. Although it is not superconducting \cite{Zhou_09}, YBaCuFeO$_{5}$ displays magnetism-driven ferroelectricity at unexpectedly high temperatures (T $<$ $T_{N2}$ $\sim$ 230K) \cite{Kundys_09} and in a temperature range more than 10 times larger than CuO. The reported  \textit{P} values are also fairly large, reaching 0.4 $\mu$C/cm$^2$  in powder samples \cite{Kundys_09}. In contrast to CuO, the ICM magnetism observed in the ferroelectric phase of YBCFO is somewhat surprising as the high symmetry tetragonal perovskite structure of this material is \textit{a priori} not prone to magnetic frustration. A further unanswered question is the nature of the ICM magnetic order, which to our best knowledge has not been reported although its existence has been known since 1995 \cite{Caignaert_95}. To understand its origin and the consequent multiferroism, a detailed knowledge of the magnetic structure and exchange interactions is clearly required.

Here we report novel neutron diffraction results that enable us for the first time to propose a model for the ICM magnetic structure of YBaCuFeO$_5$. We also present experimental evidence suggesting the existence of a particular kind of Cu/Fe chemical disorder characterized by the existence of Cu-Fe bipyramidal dimers. Density functional theory (DFT) calculations carried out for various Fe/Cu distributions provide additional support for this scenario. We also calculate the magnetic exchange coupling parameters between next and (selected) next-nearest neighbours in order to gain insight into the origin of the magnetic order and the magnetic frustration in this material. To conclude, we discuss the polarization direction and the possible origins of the multiferroicity based in the symmetry of the ICM magnetic structure and in the similar temperature dependences of the polarization and the magnetic modulation parameters.

\section{Experimental details}

The YBaCuFeO$_5$ ceramic sample used in this work was prepared by solid state synthesis. Stochiometric  amounts of Y$_2$O$_3$ (previously pre-annealed at 950$^\circ$C for 10 hours), BaCO$_3$, CuO and Fe$_2$O$_3$  were thoroughly mixed and fired at 1050$^\circ$C in air for 50h. The resulting black powder was ground and pressed into pellets which were annealed again under the same conditions. The phase purity was checked by laboratory x-ray powder diffraction (Brucker D8 Advance, Cu K$_\alpha$), which indicated the absence of foreign phases as well as an excellent crystallinity. The cationic distribution in the sample, as determined by x-ray fluorescence spectroscopy, was found to be homogeneous within a 30$\mu$m scale with Y/Ba/Fe/Cu ratios in excellent agreement with the nominal composition. The oxygen content, as determined from thermogravimetric analysis, was 4.95(2).

DC magnetic susceptibility measurements were carried out on a commercial Physical Properties Measuring System (Quantum Design). A cylinder-shaped YBaCuFeO$_5$ pellet (D = 5mm, H = 7mm) from the same batch as the sample used for the neutron diffraction measurements was cooled in zero field down to 1.8K. The magnetization of the sample was then measured in a magnetic field of 1T while heating at a constant rate of 2K/min.

The electric polarization was determined from pyrocurrent measurements which were carried out using a Keithely 6517B electrometer. A thin pellet (D = 11mm, H = 1mm) was sputtered with gold on both faces and mounted on the stick of a He cryostat. The sample was cooled from RT down to 3K with an electric field of $\pm$ 300V applied between the gold-covered faces. At 3K the field was removed and the stray charges (if any) recorded during 600s. The pyrocurrent was then measured by heating the sample at 20K/min.

Neutron powder diffraction (NPD) measurements were carried out at the Swiss Neutron Source SINQ of the Paul Scherrer Institut in Villigen. Several patterns were recorded between 1.5 and 500K at the powder diffractometers HRPT ($\lambda$=1.1546${\AA}$) \cite{HRPT_00} and DMC ($\lambda$=2.45${\AA}$) \cite{DMC_90}. The two series of experiments were carried out consecutively using the same cryofurnace, whose contribution to the background was minimized using oscillating radial collimators.  All data were analyzed using the Rietveld package FullProf Suite \cite{FullProf_93}.

\section{Crystal structure and Fe/Cu distribution}

\begin{figure*}[tbh]
\includegraphics[keepaspectratio=true,width=165 mm]{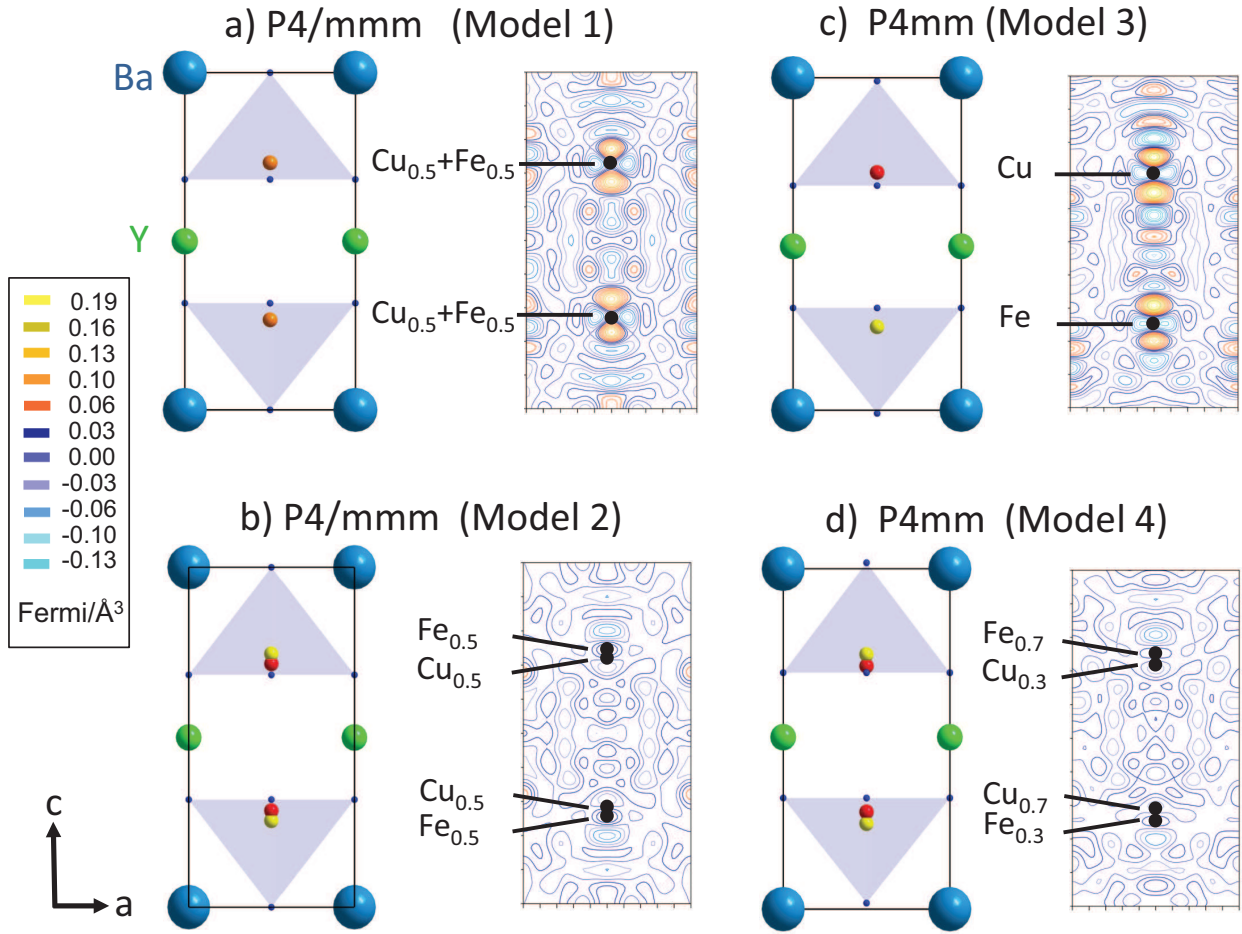}
\vspace{-3mm}
\caption{(Color online) Models for the crystal structure of YBaCuFeO$_5$ used for the Rietveld fits of the RT NPD data recorded on HRPT (see also Table ~\ref{tab:tab-structure}). (a) Centrosymmetric (P4/mmm) with a single Fe/Cu site. (b) Centrosymmetric (P4/mmm) with Fe/Cu sites split. (c) Non-centrosymmetric (P4mm) with perfect Fe/Cu order along \textbf{c}. (d) Non-centrosymmetric (P4mm) with partial Fe/Cu order. The contour plots are Fourier difference maps of the (x, 1/2, z) plane showing the neutron scattering density not reproduced by each of the models.}
\label{fig:Structure_FDmaps}\vspace{-5mm}
\end{figure*}

The crystal structure of YBaCuFeO$_5$ is displayed in Fig.~\ref{fig:Structure_FDmaps}. The tetragonal unit cell can be described as an ordered array of layers containing the large Ba$^{2+}$ ions plus two corner-sharing square pyramids separated by Y$^{3+}$ sheets. Equal amounts of Fe$^{3+}$ and Cu$^{2+}$  sit inside the pyramids, though at different distances from the basal plane \cite{Caignaert_95}. If the two ions are equally distributed among the pyramids as shown in Figs.~\ref{fig:Structure_FDmaps}a and b, the average structure is centrosymmetric with space group (SG)\textit{P4/mmm}. If the Fe/Cu distribution is asymmetric as in Figs.~\ref{fig:Structure_FDmaps}c and d, the mirror plane containing the Y$^{3+}$ ions is lost and the structure is non-centrosymmetric (SG \textit{P4mm}). Perfect Fe/Cu order along the \textbf{c} axis is a particular case of this scenario (see Fig.~\ref{fig:Structure_FDmaps}d).

Due to the relatively small difference between the Fe$^{3+}$ and Cu$^{2+}$ ionic radii (0.07{\AA}  \cite{Shannon_76}), the occupation of the two sites is in fact strongly dependent on the preparation method and the two space groups have been reported by different authors in the past \cite{Rakho_88,Atanassova_93,Mombru_94,Caignaert_95,RuizAragon_98}. This has an impact on the magnetic interactions and on the value of  $T_{N2}$, which displays an important dispersion in the literature (180 to 240K) \cite{Rakho_88,Mombru_94,Caignaert_95,Mombru_98,RuizAragon_98,Kundys_09,Kawamura_10}. A proper comprehension of the magnetism in YBaCuFeO$_5$ thus requires thus information about the Cu/Fe distribution within the 2 available square-pyramidal sites.

The results obtained for the refinement of the high resolution NPD data measured on HRPT ($\lambda$ = 1.1546 \text{\AA}) at room temperature (RT) are summarized in Tab.~\ref{tab:tab-structure}. We used two different Fe/Cu distribution models for each SG. For \textit{P4/mmm} we compared a single site occupied by 50$\%$ Fe and 50$\%$ Cu with a model with split Fe and Cu sites, each of them also with half occupation. As shown in Table~\ref{tab:tab-structure}, the fit obtained using split sites results in significantly better reliability factors.

\begin{figure}[tbh]
\includegraphics[keepaspectratio=true,width=\columnwidth]{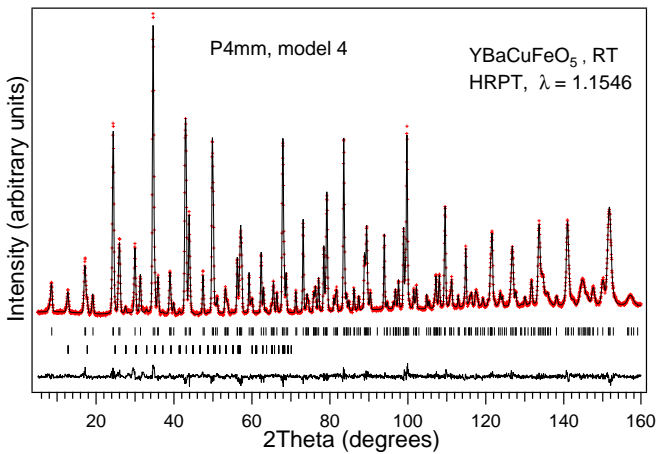}
\vspace{-3mm}
\caption{(Color online) NPD pattern measured on HRPT at RT. Red crosses: observed data. Black lines: Rietveld fit obtained using the model 4 of Table ~\ref{tab:tab-structure}. The vertical ticks indicate the positions of the Bragg reflections for the crystal (upper row) and the commensurate AFM magnetic structure with \textbf{k$_c$} = (1/2, 1/2, 1/2) (lower row, see also Fig.~\ref{fig:Magn_structures}a).}
\label{fig:Rietveld_fit_RT}\vspace{-5mm}
\end{figure}

A similar conclusion was derived from the fits using  \textit{P4mm}, which, in contrast to centrosymmetric \textit{P4/mmm}, allows us to refine the Fe/Cu occupation. For this space group we compared a model with full Fe/Cu order with another with split Fe/Cu sites. The agreement between the observed and the calculated patterns was again significantly better for the second case, which displayed the best reliability factors of the four models presented in Table~\ref{tab:tab-structure} (see also Fig.~\ref{fig:Rietveld_fit_RT}). The occupation of the split Fe/Cu sites obtained with this last model is slightly asymmetric with approximately 1/3 Fe + 2/3 Cu (\textit{z} $\sim$ 0.25) and  2/3 Fe + 1/3 Cu (\textit{z} $\sim$ 0.75). This indicates that the material is non-centrosymmetric on average with a polar axis parallel to the 4-fold axis along \textbf{\textit{c}}. As we will show later, no permanent polarization is observed above $T_{N2}$ = 200K, probably because the Fe/Cu disorder prevents the existence of coherence between the Fe/Cu displacements.

The existence of disorder is also supported by the Fourier difference maps calculated for the four models, which are displayed in Fig.~\ref{fig:Structure_FDmaps}.  We used the FullProf Suite program \textit{Fourier} \cite{FullProf_93} and structure factors corresponding

\begin{widetext}
\begin{center}
\begin{longtable}{lccccc}
\label{tab:tab-structure}
                  & \textbf{MODEL 1}   & \textbf{MODEL 2}  & \textbf{MODEL 3}      & \textbf{MODEL 4}\ \\
\textbf{T = 300K} & \textbf{\textit{P4/mmm}}   & \textbf{\textit{P4/mmm}}  & \textbf{\textit{P4mm}}      & \textbf{\textit{P4mm}}\ \\
\textbf{ }        & \textit{(1 single Fe/Cu site)} & \textit{(Fe/Cu sites split)} & \textit{(Fe/Cu fully ordered)} & \textit{(Fe/Cu split, partial order)}\ \\
\ \\
\hline
\ \\
 \textbf{a}(\text{\AA})     & 3.87323(1)             & 3.87326(1)             & 3.87325(1)                 &  3.87325(1)\ \\
 \textbf{c}(\text{\AA})     & 7.6651(3)              & 7.6655(3)              & 7.6651(3)                  &  7.6655(3)\ \\
 \ \\
 \textbf{Ba}                &\textbf{1a (0 0 0)}     &\textbf{1a (0 0 0)}     &\textbf{1a (0 0 z)}         &\textbf{1a (0 0 z)}\ \\
 \textit{z}                 &                        &                        &  0                         &  0 \ \\
 $U_{11}(\text{\AA}^2)$     & 0.0042(4)              & 0.0043(3)              &  0.0024(4)                 &  0.0035(3)\ \\
 $U_{33}(\text{\AA}^2)$     & 0.0189(12)             & 0.0231(12)             &  0.014(2)                  &  0.024(2)\ \\
 \ \\
 \textbf{Y}                 &\textbf{1b (0 0 1/2)}   &\textbf{1b (0 0 1/2)}   &\textbf{1a (0 0 z)}         &\textbf{1a (0 0 z)}\ \\
 \textit{z}                 &                        &                        &  0.4931(10)                &  0.5053(16)\ \\
 $U_{iso}(\text{\AA}^2)$    & 0.00263(17)            & 0.00250(14)            &  0.00281(19)               &  0.00249(18)\ \\
 \ \\
 \textbf{Cu}                &\textbf{2h (1/2 1/2 z)} &\textbf{2h (1/2 1/2 z)} &\textbf{1b (1/2 1/2 z)}     &\textbf{1b (1/2 1/2 z)}\ \\
 \textit{Occ}               & 0.5                    & 0.5                    &  1                         &  0.703(2) \ \\
 \textit{z}                 & 0.26729(13)            & 0.2833(3)              &  0.7155(10)                &  0.2856(7)\ \\
 $U_{iso}(\text{\AA}^2)$    & 0.00435(11)            & 0.00156(12)            &  0.00402(12)               &  0.00149(13)\ \\
 \ \\
 \textbf{Fe}                &\textbf{2h (1/2 1/2 z)} &\textbf{2h (1/2 1/2 z)} &\textbf{1b (1/2 1/2 z)}     &\textbf{1b (1/2 1/2 z)}\ \\
 \textit{Occ}               & 0.5                    & 0.5                    &  1                         &  0.297(2)   \ \\
 \textit{z}                 & 0.26729(13)            & 0.2544(2)              &  0.2511(10)                &  0.2516(8) \ \\
 $U_{iso}(\text{\AA}^2)$    & 0.00435(11)            & 0.00156(12)            &  0.00402(12)               &  0.00149(13)\ \\
 \ \\
 \textbf{O1}                &\textbf{1c (1/2 1/2 0)} &\textbf{1c (1/2 1/2 0)} &\textbf{1b (1/2 1/2 z)}     &\textbf{1b (1/2 1/2 z)}\ \\
 \textit{z}                 &                        &                        &  -0.0096(14)               &  0.0179(15)\ \\
 $U_{11}(\text{\AA}^2)$     & 0.0074(4)              & 0.0067(4)              &  0.0084(5)                 &  0.0067(4)\ \\
 $U_{33}(\text{\AA}^2)$     & 0.0136(9)              & 0.0145(8)              &  0.0157(11)                &  0.0090(17)\ \\
 \ \\
 \textbf{O2}                &\textbf{4i (0 1/2 z)}   &\textbf{4i (0 1/2 z)}   &\textbf{2c (1/2 0 z)}       &\textbf{2c (1/2 0 z)}\ \\
 \textit{z}                 & 0.31603(12)            & 0.31601(10)            &  0.3077(12)                &  0.3265(15)\ \\
 $U_{11}(\text{\AA}^2)$     & 0.0050(3)              & 0.0043(3)              &  0.0052(3)                 &  0.0043(3)\ \\
 $U_{22}(\text{\AA}^2)$     & 0.0022(3)              & 0.0024(2)              &  0.0024(3)                 &  0.0023(2)\ \\
 $U_{33}(\text{\AA}^2)$     & 0.0113(4)              & 0.0090(3)              &  0.0112(10)                &  0.0086(4)\ \\
 \ \\
 \textbf{O2'}                &                        &                        &\textbf{2c (1/2 0 z)}       &\textbf{2c (1/2 0 z)}\ \\
 \textit{z}                 &                        &                        &  0.6758(12)                &  0.6947(15)\ \\
 $U_{11}(\text{\AA}^2)$     &                        &                        &  0.0052(3)                 &  0.0043(3)\ \\
 $U_{22}(\text{\AA}^2)$     &                        &                        &  0.0024(3)                 &  0.0023(2)\ \\
 $U_{33}(\text{\AA}^2)$     &                        &                        &  0.0112(10))               &  0.0086(4)\ \\
 \ \\
 \hline
 \ \\
 $Chi^2$                    & 2.64                   & 2.01                   & 2.46                       & 1.98\ \\
 $R_p$                      & 4.52                   & 3.90                   & 4.33                       & 3.86\ \\
 $R_{wp}$                   & 5.74                   & 5.01                   & 5.54                       & 4.96\ \\
 $R_{Bragg}$                & 5.34                   & 4.05                   & 4.98                       & 3.81\ \\

 \hline
\ \\
\caption[width=175]{Atomic coordinates, thermal parameters and Fe/Cu magnetic moments of YBaCuFeO$_5$ at 300K, as refined in the space groups \textit{P4/mmm} and \textit{P4mm} (both with Z=1) using the neutron powder diffraction data recorded on HRPT ($\lambda$ = 1.1546 \text{\AA}). The reliability factors of the different models are also provided.}
\end{longtable}
\end{center}
\end{widetext}

 to reciprocal space vectors $\textbf{H}$ with modulus smaller than 10.7 {\AA}$^{-1}$. The contour plots in Fig.~\ref{fig:Structure_FDmaps} represent the Fourier transforms of the difference between the observed and the calculated neutron scattering density at RT. The results obtained for the two ordered models (1 and 3 in  Table~\ref{tab:tab-structure}) show the existence of scattering density (bright yellow spots) not reproduced by these models above and below the refined Cu/Fe positions (black circles). These spots are absent for the disordered models 2 and 4, with results slightly better for the second one.
The models involving Cu/Fe disorder (that we assume to be random in our fits) provide thus a better description of the experimental data, even if the presence of correlations between the Fe and Cu site occupations cannot be completely disregarded. In particular, it has been suggested that the bipyramidal units linked by the apical oxygen O1 could always host Cu-Fe pairs, which would be randomly arranged in the structure \cite{Caignaert_95}. Since the Fe-O1 and Cu-O1 apical distances are slightly different, such a distribution is expected to give rise to lower microstrains along \textbf{c} than one with coexisting Cu-Cu (long), Fe-Fe (short) and Cu-Fe (intermediate) pairs.  We show later that both the proposed magnetic structures and the DFT calculations favor such a scenario.

\section{Magnetic transitions}

\begin{figure}[tbh]
\includegraphics[keepaspectratio=true,width=\columnwidth]{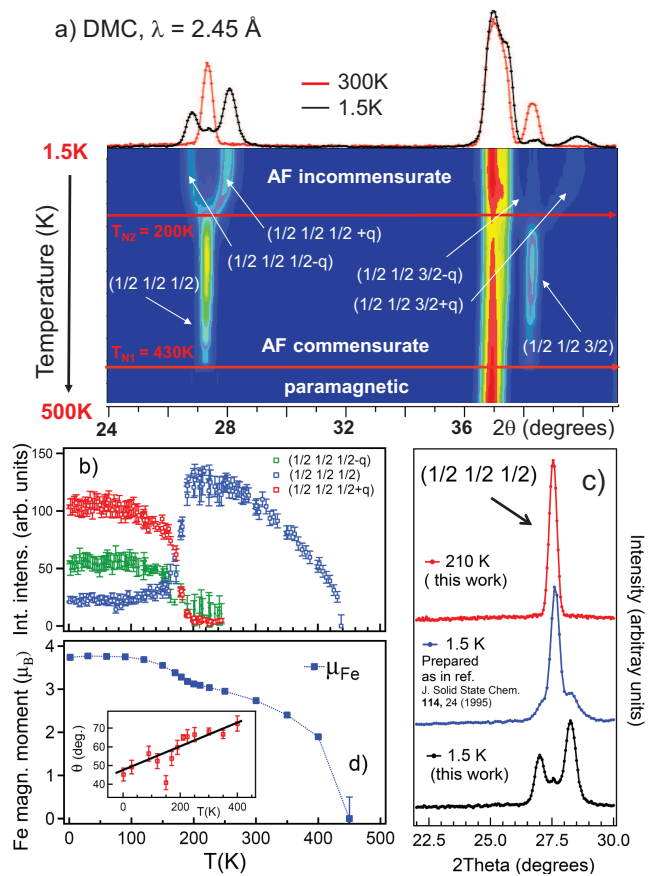}
\vspace{-3mm}
\caption{(Color online) a) 2D contour plot showing the temperature dependence of the NPD patterns for YBaCuFeO$_5$. The patterns at 1.5K and 300K are shown separately. b) Temperature dependence of the integrated intensity of the (1/2, 1/2, 1/2) magnetic reflection and its incommensurate satellites. c) Portion of the NPD patterns showing these reflections for our sample (230K and 1.5K) and a sample prepared according to ref.\cite{Caignaert_95} (1.5K). d) Temperature dependence the Fe$^{3+}$ magnetic moment. Inset: angle $\theta$ between the magnetic moment direction ($T_{N1}$$<$T$<$$T_{N2}$) and the spiral plane (T$<$$T_{N2}$) with the \textbf{c} axis.
}
\label{fig:DMC_contour}\vspace{-5mm}
\end{figure}

Fig.~\ref{fig:DMC_contour}a shows the temperature dependence of the low-angle region of the NPD patterns recorded on DMC. Two phase transitions involving the appearance of new Bragg reflections are clearly observable at $T_{N1}$ $\sim$ 440K and $T_{N2}$ $\sim$ 200K. The new peaks appearing below $T_{N1}$  correspond to the propagation vector \textbf{k$_c$} = (1/2, 1/2, 1/2), indicating that the magnetic structure is commensurate (CM) with the crystallographic unit cell between 440 and 200K in agreement with previous reports \cite{Caignaert_95,Mombru_94,Caignaert_95,Mombru_98,RuizAragon_98,Kundys_09,Kawamura_10}.

\begin{figure}[tbh]
\includegraphics[keepaspectratio=true,width=\columnwidth]{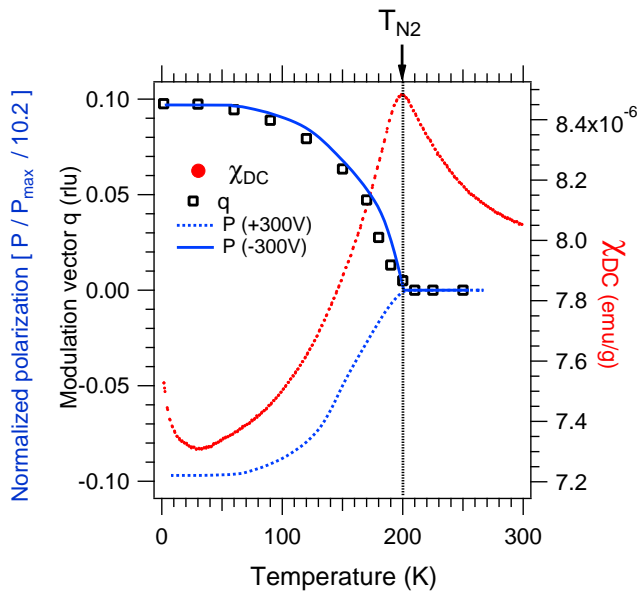}
\vspace{-3mm}
\caption{Red dots: DC magnetic susceptibility of YBaCuFeO$_5$ measured at 1T by heating after cooling in zero field. Black open squares: Incommensurate modulation vector \textit{q} (in reciprocal lattice \textbf{\textit{c$^*$ }}units). Blue dotted/continuous lines: normalized electric polarization measured by applying an electric field of $\pm$ 300V. In order to use the same axis as q the polarization values have been normalized to their saturation value ($\sim$ 0.64$\mu$C/cm$^2$) and further divided by a constant value of $\sim$ 10.2 .}
\label{fig:Chi_P}\vspace{-5mm}
\end{figure}

Below $T_{N2}$ two satellites appear around each CM magnetic reflection. They can be indexed with the propagation vector \textbf{k$_i$}=(1/2, 1/2, 1/2$\pm$\textit{q}), which involves an ICM modulation of the magnetic moments along the $\textbf{c}$ axis. The modulation parameter $\textit{q}$ increases continuously with decreasing temperature and remains ICM down to 1.5K, see Fig.~\ref{fig:Chi_P}. The onset of the ICM magnetic order coincides with a sharp anomaly in the magnetic susceptibility and with the appearance of electric polarization \textit{P} (Fig.~\ref{fig:Chi_P}), indicating a direct relationship between the two phenomena. Below 50K \textit{P} reaches $\sim$ 0.64$\mu$C/cm$^{2}$, close to the value reported by Kundys\cite{Kundys_09} ($\sim$ 0.4$\mu$C/cm$^2$) and about 10 times larger than the one reported by Kawamura\cite{Kawamura_10} (0.04$\mu$C/cm$^2$). The temperature dependence of \textit{P} closely follows that of \textit{q}, a behavior that we will address later in the text.

\smallskip
\smallskip

Fig.~\ref{fig:DMC_contour}b shows the evolution of the integrated intensities of the first magnetic peak and its satellites. Below $T_{N2}$ the main CM reflection (1/2, 1/2, 1/2) starts to decrease. At the same time the intensities of the two satellites start growing. At 1.5K the CM peak is still visible, but its intensity is much lower than that of the ICM satellites. A similar behaviour is observed for all CM/ICM reflection sets, see Figs.~\ref{fig:DMC_contour}a and ~\ref{fig:Rietveld_fit_1p5K}. Such behavior contrasts with previous reports, where the ICM satellites were either absent \cite{RuizAragon_98} or much less intense than the CM reflections at all temperatures \cite{Caignaert_95,Mombru_98,Kawamura_10}. This is illustrated in Fig.~\ref{fig:DMC_contour}c, where the region around the first CM reflection (1/2, 1/2, 1/2) measured at 1.5K for our sample (black) and a sample prepared according to the method described in ref. \cite{Caignaert_95} (blue) are displayed. The improved quality of the ICM magnetic reflection set obtained with our synthesis procedure, together with the larger number of reflections measured compared with previous works \cite{Caignaert_95,Mombru_98,Kawamura_10}, enables us for the first time to propose a model for the ICM magnetic structure of YBaCuFeO$_5$.

\section{Magnetic structures}

We used representation analysis to find the possible magnetic moment arrangements compatible with the the space groups \textit{P4/mmm} and \textit{P4mm}. The characters of the irreducible representations (\textit{irreps}) of the little group G$_\textbf{k}$, formed by the operations of the space group G which leave the \textit{\textbf{k}}-vector invariant, are shown in the Appendix A and B for both \textit{P4/mmm} and \textit{P4mm} and the two propagation vectors \textbf{k$_c$} and \textbf{k$_i$} together with tables with the symmetry-allowed basis functions associated with each \textit{irrep}. Such tables indicate that, if only one \textit{irrep} becomes active below each of the magnetic transitions, the direction of the moments is restricted to either the \textbf{ab} plane or along the \textbf{c} axis. We find instead that, at all temperatures, the best fits to the data correspond to inclined arrangements needing the combination of 2 \textit{irreps}, see the Appendix A and B for details.

\begin{figure}[tbh]
\includegraphics[keepaspectratio=true,width=\columnwidth]{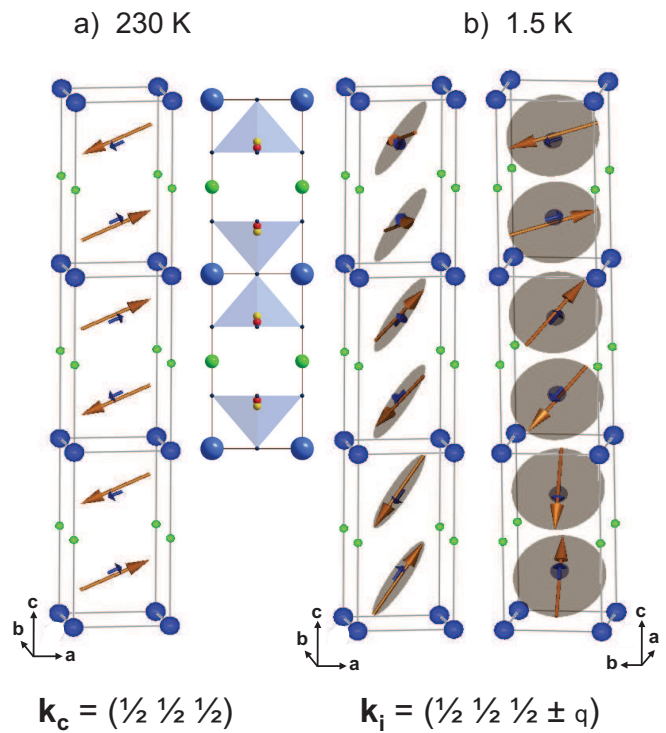}
\vspace{-3mm}
\caption{(Color online)Magnetic structures of YBaCuFeO$_5$. a) Colinear magnetic order at 230K. b) Two views of the circular spiral order at 1.5 K. For clarity, only one crystal cell within the \textbf{ab} plane is shown. }
\label{fig:Magn_structures}\vspace{-5mm}
\end{figure}

Two collinear models and a commensurate helix are compatible with our observations and they give rise to the same neutron powder diffraction patterns if Cu and Fe are located exactly at \textit{z} = 1/4 and \textit{z} = 3/4. For the refined Fe and Cu positions, which have z coordinates rather close, but not identical to these values (see Table~\ref{tab:tab-structure}) the three models give rise to slightly different intensities. As shown in the Fig.~\ref{fig:DMC_fits_collinear}, the best agreement corresponds to the collinear magnetic structure displayed in Fig.~\ref{fig:Magn_structures}a. Spins in the \textbf{ab} plane are antiferromagnetically (AFM) coupled whereas the alignment along $\mathbf{c}$ alternates: it is AFM across the O-free Y planes and ferromagnetic (FM) inside of the bipyramidal blocks. Note that the orientation of the spins the \textbf{\textit{ab}} plane is arbitrary because their direction in the plane perpendicular to the 4-fold axis cannot be determined from NPD. The angle $\theta$ between magnetic moments and $\mathbf{c}$ axis decreases continuously from 75$^{\circ}$ (at 400K) to 65$^{\circ}$ (at 200K), see inset of Fig.~\ref{fig:DMC_contour}d.

\begin{figure}[tbh]
\includegraphics[keepaspectratio=true,width=\columnwidth]{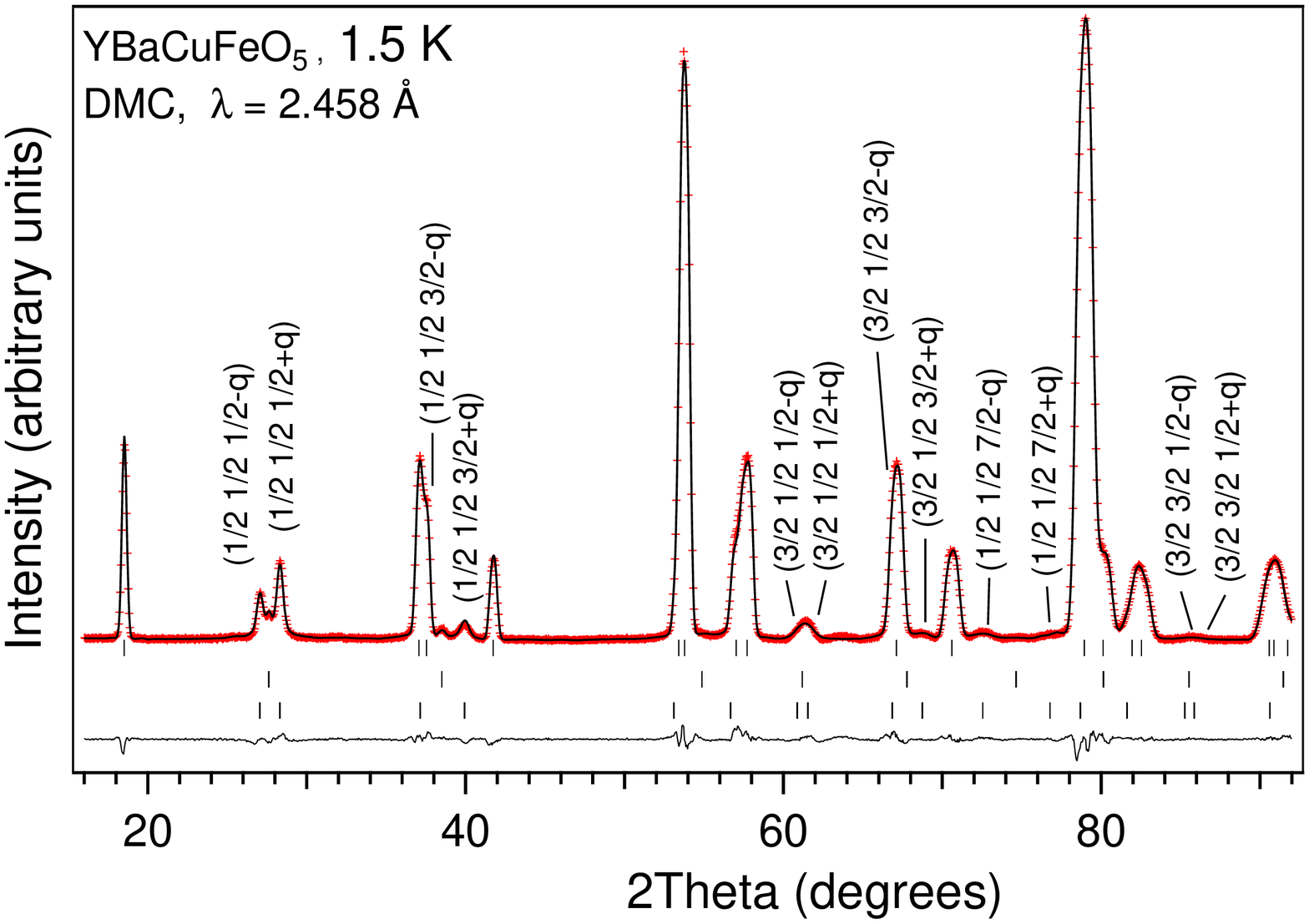}
\vspace{-3mm}
\caption{(Color online) NPD pattern measured on DMC at 1.5K. Red crosses: observed data. Black lines: Calculated intensities. The vertical ticks indicate the positions of the Bragg reflections for the crystal (upper row), the CM magnetic structure with \textbf{k$_c$} = (1/2, 1/2, 1/2) (intermediate row) and the ICM magnetic structure with \textbf{k$_c$} = (1/2, 1/2, 1/2$\mp$\textit{q}) (lower row).}
\label{fig:Rietveld_fit_1p5K}\vspace{-5mm}
\end{figure}

Below $T_{N2}$ two propagation vectors \textbf{k$_c$} and \textbf{k$_i$} are present down to the lowest temperature investigated, see Fig.~\ref{fig:DMC_contour}b and Fig.~\ref{fig:Rietveld_fit_1p5K}. The magnetic intensities can be described as arising from either a multi-\textit{\textbf{k}} arrangement or from two distinct magnetic phases; our NPD data alone cannot distinguish between the two possibilities. However, the fact that the intensities of both reflection sets have opposite temperature dependencies below $T_{N2}$ favors the second one.

As in the CM case, two different models give rise to very similar magnetic intensities. However, the excellent statistics of the DMC data and the fact that Fe and Cu z coordinates differ significantly from \textit{z} = 0.25 and 0.75 at all temperatures allow us to distinguish them, see Fig.~\ref{fig:Rietveld_fit_1p5K}, Fig.~\ref{Fig:DMC_fits_spiral} and further details in the Appendix B. The best agreement with the experimental data corresponds to the circular helix displayed in Fig.~\ref{fig:Magn_structures}b. Such a spin arrangement involves the preservation of the AFM coupling between the 3d metal sites without connecting oxygen observed in the commensurate phase, see Fig.~\ref{fig:Magn_structures}a. It also implies the loss of the FM coupling within the bipyramids, suggesting that this magnetic coupling is more affected by the thermal evolution of the structure than is the one across the O-free Y layers.

Our results indicate that the plane of the helix forms an angle $\theta$ with the \textbf{c} axis which decreases continuously with temperature ($\theta$ $\sim$ 65$^\circ$ at 200K, $\theta$ $\sim$ 45$^\circ$ at 1.5K, the anomalously low value at 150K is probably due to the strong superposition of the 2 incommensurate satellites at this temperature). The ICM magnetic order evolves thus from an inclined helix towards a cycloid with decreasing temperature. The reasons for this behaviour are unclear, but it could be related to different Cu$^{2+}$ and Fe$^{3+}$ magnetic anisotropies and their effect on the temperature dependence of the magnetic moment orientations.

Using the previously described models for the CM and ICM magnetic structures we determined their fractions in the sample below $T_{N2}$. By restricting the modulus and the $\theta$ angle of the Cu$^{2+}$ and Fe$^{3+}$ magnetic moments in both phases to be identical and their ratio to be the same as for their spin-only values (1 to 5) we obtain 8$\%$ (CM) and 92 $\%$ (ICM). This contrasts with previous studies, where most of the sample displayed CM magnetic order below $T_{N2}$ and only a small fraction was able to undergo the CM $\rightarrow$ ICM phase transition. As we discuss in the next sections, differences in the Fe/Cu distribution are the most probable origin of these differences.

The evolution of the Fe$^{3+}$ magnetic moment is shown in Fig.~\ref{fig:DMC_contour}d. Its value at 1.5 K is 3.74(2) (0.748(4)$\mu_B$ for Cu$^{2+}$), about 1/3 reduced with respect to those expected for the free-ion, spin-only moments. Such reduction may, among other reasons, also be related to Fe/Cu disorder. The dip around $T_{N2}$ is not an artifact from the fits since the same dip is observed in the total (central peak + satellites) integrated intensity as a function of temperature (Caignaert and co-workers found a similar behaviour, see \cite{Caignaert_95}). This suggests that the incommensurate domains are large enough to produce a mesurable contribution to satellites only a few degrees below $T_{N2}$.

\section{Magnetic couplings from Goodenough-Kanamori-Anderson rules}

To understand the origin of the magnetic structures we first consider the Goodenough-Kanamori-Anderson (GKA) rules of superexchange. For Fe$^{3+}$ (3d$^{5}$) and Cu$^{2+}$ (3d$^{9}$) in the square-pyramidal arrangement of the YBaCuFeO$_5$ structure, these rules predict strong AFM nearest-neighbour (NN) couplings within the \textbf{ab} plane in agreement with our observations. Interestingly, \textit{the sign of the couplings is independent of the Cu/Fe distribution}. Along the \textbf{c} axis the situation is more complex. On one side, superexchange between NN is not possible across the Y layers due to the absence of the apical O. On the other, the sign of the coupling within the bipyramids is expected to be strongly dependent on the Fe/Cu distribution (AFM for two Fe ions, FM for a Fe-Cu pair and negligibly small for two Cu ions). This agrees with the observation of an ICM modulation \textit{only} along this direction and suggests that the appearance of magnetic frustration below $T_{N2}$ could be due to a temperature-driven imbalance of the magnetic couplings along the \textbf{c} axis. Such a scenario is consistent with the continuous evolution of the modulation parameter \textit{q} with decreasing temperature (Fig.~\ref{fig:Chi_P}). Note also that, within the GKA framework and for the CM phase, the experimentally observed FM coupling within the bipyramids \textit{is only possible if they are occupied by a Cu-Fe pair}.

\section{Ab initio Calculations}
Further insight requires detailed quantitative calculations of the exchange interactions that we obtain using {\it ab initio} calculations with the Local Spin Density Approximation plus Hubbard U (LSDA+U)\cite{Perdew1981} to density functional theory as implemented in the Vienna ab-initio simulation package (VASP) \cite{Kresse1,Kresse2}.
For these calculations we used projector augmented wave potentials with 3$d$ and 3$p$ orbitals treated as valence states for both Fe and Cu ions.
The values of on-site effective Coulomb interaction were set to U$_{\text{Fe}}=5$~eV and U$_{\text{Cu}}=8$~eV while the on-site effective Hund's couplings were J$^{H}_{\text{Fe}}=1$~eV and J$^{H}_{\text{Cu}}=0$ for Fe and Cu, respectively.
For all the calculations, except for those relative to some of the next-nearest-neighbor magnetic couplings along $\mathbf{c}$ (see Appendix C), we considered a supercell with \textbf{a} = $\surd$2 \textbf{a$_c$} and \textbf{c} = 2\textbf{c$_c$} (\textbf{a$_c$} and \textbf{c$_c$} are the crystallographic unit cell parameters) containing four formula units, as depicted in Fig.~\ref{fig:Structures}, a $\Gamma$-centered k-point grid of size $8\times 8 \times 4$ and an energy cut-off E$_{\text{cut}}=600$~eV.
Relaxations of the reduced ionic positions in the unit cell were performed until all atomic forces were below $2\cdot 10^{-5}$ eV/{\AA} and using the experimental lattice parameters at 1.5K (\textbf{a}=\textbf{b}=5.462 \AA,  \textbf{c}=15.258 \AA~for the supercell.).

\subsection{Cu/Fe distribution}
\begin{figure}[tbh]
\includegraphics[keepaspectratio=true,width=75 mm]{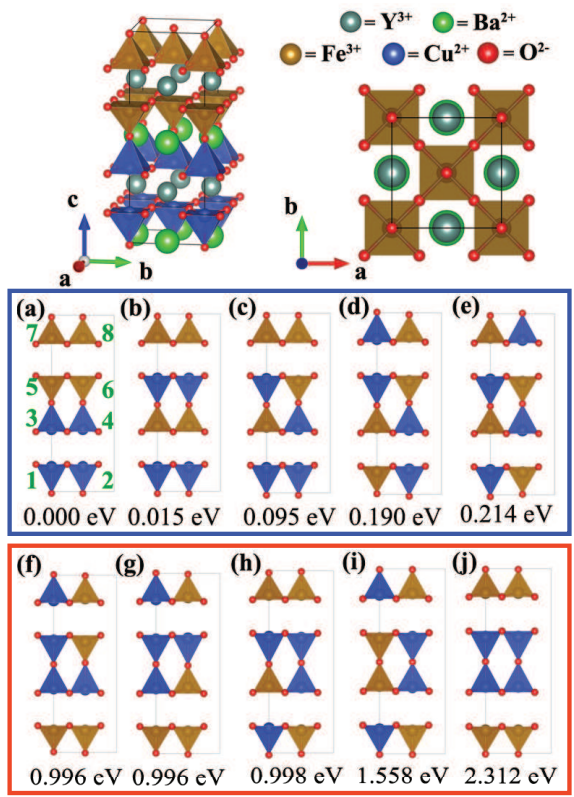}
\vspace{-3mm}
\caption{ (Color online) Three-dimensional view of the supercell with \textbf{a} = $\surd$2 \textbf{a$_c$} and \textbf{c} = 2\textbf{c$_c$} used in the \textit{ab-initio} calculations (left upper panel) and its projection on the $\mathbf{ab}$ plane (right upper panel) for one of the Fe$^{2+}$/Cu$^{3+}$ distributions considered.
Note that the $\mathbf{a}$ and $\mathbf{b}$ axes are rotated by $\frac{\pi}{4}$ with respect to those in Fig.~\ref{fig:Structure_FDmaps}.
Panels from (a) to (j) show the projection on the $\mathbf{ac}$ plane of the considered supercells with different orderings of Fe$^{2+}$ ions (golden square pyramids) and Cu$^{3+}$ (blue square pyramids)  and their energies relative to that of structure (a). For clarity Ba$^{2+}$ and Y$^{3+}$ ions are not shown. The blue and red frames surround respectively the low-energy and high-energy sets mentioned in section VII.A.}
\label{fig:Structures}\vspace{-5mm}
\end{figure}

First, we investigated the stability of the crystal structure of YBaCuFeO$_5$ under different Fe$^{3+}$/Cu$^{2+}$ distributions. The fractional coordinates of all atoms were relaxed starting from different arrangements for the two transition metal ions, keeping the commensurate antiferromagnetic order of Fig.~\ref{fig:Magn_structures}a, fixing the lattice parameters to the experimentally observed values at 1.5K and in absence of the spin-orbit coupling. Fig.~\ref{fig:Structures} shows the different configurations for the Fe/Cu distribution and the energy of the relaxed structures compared with the lowest energy one. We see that there are two energy hierarchies depending on which ions are present in the bipyramids. Low energy configurations (up to $0.214$ eV $\approx$ $2500$ K), panels (a) to (e) within the blue frame, contain \textit{both} Fe$^{3+}$ and Cu$^{2+}$ in all bipyramids, while those containing two Cu$^{2+}$ and/or two Fe$^{3+}$ ions in at least one of the bipyramids have higher energy (from $0.996$ eV $\approx$ $11600$ K), panels  (f) to (j) within the red frame .\\

\subsection{Nearest-neighbour magnetic couplings and CM magnetic order}
Next, we evaluated the exchange coupling constants for the obtained relaxed structures of the low energy configurations (a) to (e), i.e., those where the bipyramids are always occupied by a Fe-Cu pair. We assumed that the magnetic interactions are described by the Heisenberg Hamiltonian
\begin{equation}
 H = \frac{1}{2}\sum_{i,j} J_{ij} \mathbf{S}_i \cdot \mathbf{S}_j,
\label{Eq_Heisenberg}
\end{equation}
where, $S_{\text{Fe}}=5/2$, $S_{\text{Cu}}=1/2$ and $i,j$ label the magnetic sites, and performed collinear spin polarized calculations in the absence of spin-orbit coupling to extract the exchange couplings as described in Appendix C.
We note that in this minimal model we assume the presence of only nearest-neighbour (NN) Heisenberg exchanges and do not consider higher order terms such as biquadratic and ring exchange which have been found to play a role in cuprates
\cite{Coldea2001,Calzado2003} and, more recently, in manganites \cite{Fedorova2014}.

Results for the exchange coupling constants are listed in Tab.~\ref{tabexchanges}. Remarkably, for any pair of NN, the sign of the magnetic interactions is the same for all five configurations. The exchange constant, $J_{\perp}$, between neighboring spins in the same tetragonal plane is AFM and is by far the strongest coupling for all of configurations. J$_{\parallel \mbox{O}}$ and J$_{\parallel}$, which correspond to the exchange couplings between ions in neighboring planes separated by an oxygen layer or not are, respectively, FM and AFM and, generally, take values smaller than  $J_{\perp}$. This confirms the \textbf{c} axis as the direction with the weakest couplings, which is the direction where an incommensurate wave vector is observed below $T_{N2}$. Interestingly, the only FM exchange constant is J$_{\parallel \mbox{O}}$, which corresponds to the coupling within the bipyramids occupied by a Cu-Fe pair.

We note that the signs of \textit{all} couplings in configurations (a) to (e) are in agreement with the Goodenough-Kanamori-Anderson rules as well as with the observed CM magnetic structure ($T_{N1}$ $<$ T $<$ $T_{N2}$), see Fig.~\ref{fig:Magn_structures}a. This is not the case for the higher energy configurations (f) to (j), where the sign of J$_{\parallel \mbox{O}}$ for bipyramids occupied by an Fe-Fe pair is AFM and hence not compatible with the observed magnetic order \cite{SM}. To summarize, the configurations containing exclusively Cu-Fe pairs in the bipyramids are more stable, and the coupling within such dimers is the only FM one among NN interactions. Since, as shown in Fig.~\ref{fig:DMC_fits_collinear}, the magnetic structure which gives the best reliability indexes involves FM couplings within the bipyramids, the only possibility is that such units are occupied by Cu-Fe pairs. This result strongly supports the existence of Cu-Fe ``dimers'' as necessary condition to stabilize the observed CM magnetic structure. Moreover, to be consistent with the results displayed in Fig.~\ref{fig:Structure_FDmaps}, these dimers should be randomly distributed.

\begin{table}
\centering
\begin{tabular}{|c|c|c|c|c|c }
\hline
 & J$_{\perp}$ (meV) & J$_{\parallel}$ (meV) & J$_{\parallel \mbox{O}}$ (meV)   & J$_{NNN}$ (meV)
\\
[0.4ex] \hline
(a) & {\color{blue}J$_{1,2}$ = 134.5} & {\color{blue}J$_{1,3}$ = 10.6} &J$_{5,3}$ = -1.6  & J$_{1,5}$ = -0.05 \\
  &  {\color{Tan}J$_{7,8}$ = 8.7}  & {\color{Tan}J$_{5,7}$ = 2.8}  &   & J$_{1+c,5} =$ -0.01 \\
[0.2ex]
\hline
(b) & {\color{blue}J$_{1,2}$ = 129.9} &  J$_{1,3}$ = 1.4  &  J$_{3,5}$ = -1.6 & {\color{blue}J$_{1,5}$ = 0.07}\\
 &  {\color{Tan}J$_{3,4}$ = 8.9}  &      &     &  {\color{Tan}J$_{3,7}$ = 0.19}\\
[0.2ex]
\hline
(c) & J$_{3,4}$ = 28.6 & J$_{5,7}$ = 1.1 & J$_{2,8}$ = -1.5 & {\color{Tan} J$_{3,7}$ = 0.20} \\
 & {\color{blue}J$_{1,2}$ =  133.0} & {\color{blue}J$_{2,4}$ = 8.9 }     &  J$_{1,7}$ = -1.5    & {\color{blue} J$_{1,5}$ = 0.09 } \\
 & {\color{Tan}J$_{7,8}$ = 8.7  } & J$_{1,3}$ = 1.7 & J$_{3,5}$ = -1.7  & J$_{2,6}$ = - 0.07\\
  & J$_{5,6}$ = 28.2   &  {\color{Tan}J$_{6,8}$ = 3.0 }  &  J$_{6,4}$ = - 1.7 &  J$_{2+c,6}$ = -0.01\\
[0.2ex]
\hline
(d) & J$_{1,2}$ = 28.3 & {\color{Tan}J$_{1,3}$ =  3.1} & J$_{3,5}$ = -1.6  & J$_{1,5}$ = -0.04\\
  &  & {\color{blue}J$_{7,5}$ = 7.5} & & J$_{1+c,5}$=-0.01\\
[0.2ex]
\hline
(e) & J$_{1,2}$ = 28.3  & J$_{1,3}$ = 1.3  & J$_{3,5}$ = -1.6  & {\color{blue}J$_{1,5}$ = 0.12} \\
 & & & &  {\color{Tan}J$_{2,6}$ = 0.20}\\
[0.2ex]
\hline
\end{tabular}
\caption{ (Color online) Calculated exchange coupling constants for the different configurations in Fig.~\ref{fig:Structures}. Positive/negative signs correspond to AFM/FM interactions, respectively. J$_{\perp}$ , J$_{\parallel O}$, J$_{\parallel}$ and J$_{NNN}$ are defined in the main text. To avoid ambiguities the exchange coupling constants $J_{i,j}$ are also labeled by the position $i,j$ of the magnetic ions as in Fig.~\ref{fig:Structures}. Note that according to
Eq. \ref{Eq_Heisenberg}  the energy scale for each coupling is obtained by multiplying it by $S_i S_j$, that is: $\frac{25}{4}$ for Fe$^{3+}$-Fe$^{3+}$ coupling (brown),  $\frac{5}{4} $ for Fe$^{3+}$-Cu$^{2+}$ coupling (black) and $\frac{1}{4}$ for Cu$^{2+}$-Cu$^{2+}$ coupling (blue).}
\label{tabexchanges}
\end{table}

\subsection{Next-nearest-neighbour magnetic couplings and ICM magnetic order}

We will focus now on the ICM magnetic structure observed below $T_{N2}$. To obtain further insight on its origin we considered the effect of next-nearest-neighbor $J_{NNN}$ magnetic couplings along the $\mathbf{c}$ direction. The values obtained by {\it ab initio} calculations are summarized in the fourth column of Tab.~\ref{tabexchanges}. For each of the configurations (a), (b), (d) and (f) in Fig.~\ref{fig:Structures} there are only two inequivalent $J_{NNN}$'s. For configuration (c) their number is much larger (eigth). Therefore for this configuration we extract only J$_{3,7}$, J$_{1,5}$, J$_{2,6}$ and J$_{2+c,6}$ and assume that J$_{3+c,7} \approx$J$_{3,7}$, J$_{1+c,5} \approx$J$_{1,5}$,  J$_{4,8} \approx$J$_{2+c,6}$ and  J$_{4+c,8} \approx$J$_{2,6}$.

\begin{figure}[!t]
\includegraphics[keepaspectratio=true,width=85 mm]{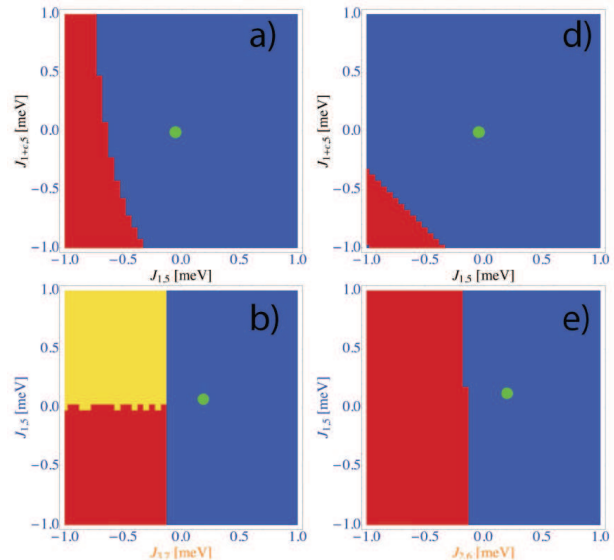}
\vspace{-3mm}
\caption{(Color online) Phase diagrams obtained using the spiral ansatz Eq. \ref{eq_SAnsatz} for configurations (a), (b), (d) and (e). The blue color indicates the commensurate collinear structure shown in the left pannel of Fig.~3.  The red color and the yellow color indicate, respectively, an incommensurate spiral state and a commensurate non-collinear state. The green dot indicates the values of the NNN in Tab.~1.}
\label{fig:PD}\vspace{-5mm}
\end{figure}

To estimate the size of $J_{NNN}$ necessary to give rise to a magnetic spiral state we calculated the phase diagrams for the minimal energy state of Eq.~\ref{Eq_Heisenberg} by tuning the inequivalent J$_{NNN}$'s of each configuration between -1 and 1 meV and under the assumption that the magnetic order is of spiral type. Calculations for configuration (c) were not considered due to the large number of inequivalent next-nearest-neighbor couplings along $\mathbf{c}$.  We assume the magnetic structure to be described by a spiral ansatz  of the form

\begin{equation}
\mathbf{S}_{\mu} (\mathbf{R}_i) = \cos (\phi_{\mu} + \mathbf{q} \cdot \mathbf{R}_i) \mathbf{v_1} + \sin (\phi_{\mu} + \mathbf{q} \cdot \mathbf{R}_i) \mathbf{v_2},
\label{eq_SAnsatz}
\end{equation}

where $\mathbf{v}_1$ and $\mathbf{v}_2$ are two orthogonal unit vectors, $\mu=1,...,8$ labels the magnetic ions in the unit cells of Fig.~\ref{fig:Structures}, $q$ is the wave vector, $\phi_\mu$ is the  angle of the $\mu$-th spin inside the unit cell and $\mathbf{R}_i$ is the position of the $i$-th unit cell. Inserting Eq. \ref{eq_SAnsatz} in Eq. 1, considering the exchange of the first three columns of Tab.~\ref{tabexchanges} and minimizing numerically the obtained energy with respect to $\mathbf{q}$ and $\phi_i$ we obtained the states depicted in the phase diagrams Fig.~\ref{fig:PD} as a function of inequivalent J$_{NNN}$.

As shown in Fig~\ref{fig:PD}, incommensurate states are stable only if at least one of the NNN couplings is ferromagnetic (i.e., if it frustrates the collinear commensurate order). According to Tab.~\ref{tabexchanges}, the only ferromagnetic NNN couplings are those between a Cu$^{2+}$ and a Fe$^{3+}$ which are present in configuration a) and d). However, the calculated strength of these couplings (green dots) is too small to stabilize a spiral state.
Therefore, the phase diagrams in Fig.~\ref{fig:PD} obtained with the ansatz Eq.~\ref{eq_SAnsatz}  indicate that the next-nearest-neighbor couplings obtained in Tab.~\ref{tabexchanges} are either too weak or of the wrong sign to  stabilize a spiral state. However, due to the uncertainty present in the ab-initio calculations of exchange couplings, we cannot completely exclude NNN coupling as the origin of the incommensurate state.

\section{Possible origin of electric polarization}

To finish, we briefly address the questions of the direction of the electric polarization in the incommensurate phase and the origin of the magnetoelectric coupling. Although they can not be properly answered due
to the lack of single crystals, it is possible to make some predictions using the symmetry properties of the refined ICM magnetic structure and the various mechanisms proposed in the literature.

Looking at Fig.~\ref{fig:Magn_structures}b, we see that the only symmetry element of the depicted inclined spiral is a binary axis along the \textit{\textbf{b}} direction. Its point group is thus .2., which is compatible with the existence of \textit{\textbf{P}} only along \textit{\textbf{b}}. Interestingly, the ionic, non-switchable polarization expected in the case of a perfect Cu/Fe order would appear along \textbf{\textit{c}}.

This prediction agrees with those of the Landau Theory formalisms proposed in Ref.~\onlinecite{Mostovoy2006} as well as in Ref.~\onlinecite{Harris_07}, where the magnetoelectric coupling is described by a trilinear interaction term. According to the formalism in Ref.~\onlinecite{Harris_07}, \textit{\textbf{P}} transforms as the product of the two \textit{irreps} describing the magnetic structure. Looking at Tables ~\ref{tab:tab-charcomm} and ~\ref{tab:tab-charincomm}, it is easy to see that \textit{\textbf{P}} is not allowed for \textit{\textbf{k$_c$ }} but it may exist in the ICM phase only within the \textbf{ab} plane.

Similar conclusions can be derived from the Dzyaloshinskii-Moriya (DM) and/or spin-current models for magnetoelectric coupling. According to these mechanisms, the polarization direction is expected to be given either by \textit{\textbf{e$_{ij}$}} $\times$ {\textbf{S$_i$} $\times$ \textbf{S$_j$} or \textit{\textbf{q}} $\times$  {\textbf{S$_i$} $\times$ \textbf{S$_j$}, were \textit{\textbf{e$_{ij}$}} is the vector connecting the \textit{i} and \textit{j} sites and \textit{\textbf{q}} = (0, 0, \textit{\textbf{k$_z$}}) is the magnetic modulation vector. In the incommensurate phase  \textit{\textbf{q}} $\parallel$ \textit{\textbf{e$_{ij}$}}, so the two models make identical predictions. From Fig.~\ref{fig:Magn_structures}b, we see that in both cases the polarization is expected to be along \textit{\textbf{b}}, in agreement with the conclusions derived from symmetry arguments.

We also speculate that the similar temperature dependence of \textit{P} and \textit{q} shown in Fig.~\ref{fig:Chi_P} might indicate that Dzyaloshinskii-Moriya (DM) and/or spin-current mechanisms could be responsible for the ferroelectricity. This may look surprising in view of the large value of the polarization in YBaCuFeO$_5$ (0.64 $\sim$ $\mu$C/cm$^2$). It is however not unreasonable since TbMnO$_3$, where multiferroicity is believed to originate from these mechanisms, displays a polarization which is only 6 times smaller ( $\sim$ 0.09 $\mu$C/cm$^2$).

From the above mentioned mechanisms, one would expect $\textit{\textbf{P}} \propto \textbf{c} \times \mathbf{S}_i \times \mathbf{S}_j$, where $i$ and $j$ are nearest-neighbor magnetic sites along \textit{$\mathbf{c}$}. As shown in Fig.~\ref{fig:DMC_contour}d, the size of magnetic moments below $T_{N2}$ is approximately constant. Therefore, one can assume the magnetic spiral state described by Eq.~\ref{eq_SAnsatz}, where $\mu=1,2$ labels the magnetic sites in Fig.~\ref{fig:Structure_FDmaps}, $\mathbf{v}_1 =\sin(\theta) \hat{\mathbf{a}} + \cos(\theta) \hat{\mathbf{c}}$, $\mathbf{v}_2 =\sin(\theta) \hat{\mathbf{b}}$ and the spin value is constant. Moreover, as the magnetic moments of neighboring magnetic ions along $z$ not separated by oxygen are antiparallel (see Fig.~\ref{fig:Magn_structures}b  and Table~\ref{tabexchanges} ), one can assume $\phi_2=\phi_1 + \pi$. Under these assumptions the electric polarization given by DM and/or spin-current coupling is $P_b = C \cos(\theta) \sin(2\pi (\frac{1}{2}-q))$ where $C$ is a temperature-independent constant.

Fig.~\ref{fig:PvsQ} shows the comparison between the measured values of $\mathbf{P}$ (blue dots) and those obtained from the previous expression setting $C$ in such a way that at the lower temperature the calculated polarization coincides with the observed value. The points represented by black dots were obtained using the measured values of $q$ and $\theta$ while for the points represented by red diamonds the values of the linear interpolation of $\theta$ below $190$ K were used (see inset of Fig.~\ref{fig:PvsQ}). Taking into account that the only free parameter is the proportionality constant $C$, the agreement between the observed and calculated temperature dependence of $\mathbf{P}$ is reasonably satisfactory.

\begin{figure}[!t]
\includegraphics[keepaspectratio=true,width=85 mm]{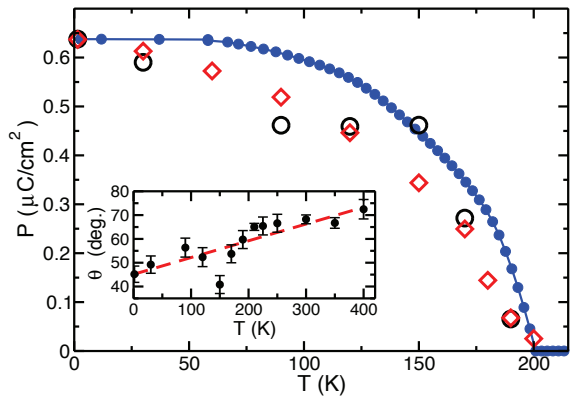}
\vspace{-3mm}
\caption{(Color online) Temperature dependence of the electric polarization (blue dots) and estimations using $\theta$ and $q$ as described in the text (red diamonds and black open dots).}
\label{fig:PvsQ}\vspace{-5mm}
\end{figure}

According to the scenario described above, the approximate proportionality of $\mathbf{P}$ and $q$ originates from the relatively small changes in $q$ (which implies $P_b\propto2 \pi C q \cos(\theta)$) and the fact that $\theta$ does not change dramatically below $T_{N2}$. We also note that this behavior would not hold if the size of magnetic moment would strongly depend on temperature below $T_{N2}$, e.g., if the magnetic spiral phase would appear close to $T_{N1}$.

\section{Summary and conclusions}

As mentioned in the introduction, YBaCuFeO$_5$ is together with CuO the only known material to display switchable, magnetism-driven ferroelectricity above 200K. A prerequisite to understand the existence of multiferroicity at such high temperatures in YBaCuFeO$_5$ is the knowledge of its ICM magnetic structure. However, it was never reported even if its existence is known since 1995. In this study we have successfully synthesized YBaCuFeO$_5$ ceramic samples of unprecedented quality and we have conducted new, high-resolution and high-intensity neutron diffraction measurements that enabled us for the first time to solve the incommensurate magnetic structure of YBaCuFeO$_5$. Our results are consistent with the replacement of the collinear magnetic order existing between $T_{N2}$  and $T_{N1}$ = 340K by a circular spiral with temperature dependent inclination described by the propagation vector \textbf{k$_i$}=(1/2, 1/2, 1/2$\pm$\textit{q}).

The origin of the ferroelectricity observed below $T_{N2}$ and its coupling with the incommensurate magnetic order can not be fully explained from our results alone, in particular because our ceramic samples do not allow to determine the direction of the polarization.  However, the symmetry of the observed magnetic spiral suggest that \textit{P} should be within the \textbf{ab} plane. Also, the very similar temperature dependence of the magnetic modulation parameter \textit{q} and the polarization indicate that the Dzyaloshinskii-Moriya and/or spin-current mechanisms could be at the origin of the magnetoelectric coupling. Further experimental work, preferably on single crystals, will be necessary to get additional insight about the validity of this scenario.

We have also investigated the crystal structure, in particular the Fe/Cu distribution between the two available square-pyramidal sites, and we find clear evidence for the existence of occupational disorder. The observed CM magnetic structure suggests however that the bipyramidal units may be preferentially occupied by Cu-Fe FM pairs.  This finding is supported by \textit{ab initio} calculations, which suggest that Cu/Fe distributions containing exclusively Cu-Fe dimers have lower ground-state energies than those where Cu-Fe, Fe-Fe and/or Cu-Cu pairs coexist. Moreover, the exchange parameters calculated for different Fe/Cu distributions are compatible with the observed CM magnetic order \textit{only} if the bipyramidal units are occupied by Cu-Fe pairs.

To summarize, all our experimental and theoretical results converge towards the same scenario, i.e., that the existence of Cu-Fe dimers may be a necessary condition for the existence of the CM magnetic structure. Elastic diffuse scattering, which provides information about occupational correlations, could be used as a further check if single crystals become available in the future.

As in the case of CuO \cite{Kimura_08}, the observation of incommensurate magnetic order below temperatures as high as 200K is most probably related to the large values of some of the exchange constants. However, in our case, it is also linked to the presence of a crystallographic direction where the magnetic couplings are particularly unstable. As seen in Fig. 5a, the CM magnetic order can be seen as an array of bilayers containing weakly coupled Cu-Fe FM bipyramidal dimers coupled through weak AFM interactions along the \textbf{c} axis. These couplings might be frustrated by several mechanisms such as NNN interactions, bond disorder or higher order magnetic couplings which could eventually give rise to a magnetic spiral. We have briefly investigated the first possibility and we find that the calculated NNN magnetic exchanges along the \textbf{c } axis are either too small or have the wrong sign to stabilize a spiral state. Although we do not completely exclude this mechanism due to the uncertainty in the \textit{ab-initio} calculations for small exchange couplings, investigation of the two remaining options is presently in progress.

The calculated exchanges highlight the strong dependence of the magnetic interactions on the Cu/Fe distribution. This could explain the huge dispersion in the $T_{N1}$ and $T_{N2}$ values reported in previous studies, where different preparation synthesis procedures were used. In general, methods favouring an intimate mixing of the 4 involved cations seem to result in higher $T_{N2}$ and larger fractions of incommensurate phase below $T_{N2}$. Although we don't have any experimental proof, we speculate that such methods may enable the material to reach more easily the most energetically favourable Cu/Fe distribution, i.e., that with bipyramids populated exclusively by Cu-Fe dimers. Because in this case all couplings along the \textbf{c} axis are weak, it may be easier to frustrate them and to generate incommensurate magnetic order. Systematic investigations of this matter are currently in progress.

As a concluding remark, we would like to point out that other perovskite oxides have been reported to display the same crystal structure as YBaCuFeO$_5$. However, the incommensurate magnetic order and the concomitant spontaneous polarization have been only found in two other members of the RBaCuFeO$_5$ family (R = Tm and Lu)\cite{Kawamura_10}. The reasons behind this are not yet fully understood, but they probably stem from the particular combination of exchange constants associated to the presence of Fe$^{3+}$ and Cu$^{3+}$, especially to the weak, alternating FM and AFM couplings along \textbf{c} axis. Since the magnetic order is particularly unstable along this direction, the application of external perturbations may provide further insight about the hierarchy of the involved magnetic interactions. Tuning such weak couplings along \textbf{\textit{c}} (and hence $T_{N2}$) could be achieved either by chemical substitution, by the application of external pressure or, in view if the layered nature of YBaCuFeO$_5$, through the fabrication of heterostructures. Given the high value of $T_{N2}$, this qualifies YBaCuFeO$_5$ as one of the most promising structural frameworks to search for room temperature multiferroics.

\section{Appendix}
The possible magnetic structures compatible with the space groups $P4/mmm$ and $P4mm$ were determined using representation analysis. The only magnetic ions in YBaCuFeO$_5$ are Fe$^{3+}$ and Cu$^{2+}$, and there is only one Fe and one Cu atom per crystallographic unit cell. The labeling of the different sites for both space groups is displayed in Tab.~\ref{tab:tab-FeCupositions}. The main difference between $P4/mmm$ and $P4mm$ is that the sites 1 (\textit{z} $\sim$ 0.25) and 2 (\textit{z} $\sim$ 0.75) are symmetry-related in the first one whereas they are not in the second. Since for a given space group Cu and Fe occupy identical Wyckoff positions, we discuss only the Fe case in the following. All conclusions apply also to Cu.

\begin{table}[ht!]
\caption{\label{tab:tab-FeCupositions}Atomic positions of Fe and Cu within the unit cell in the space groups \textit{P4/mmm} (site 2h) and \textit{P4mm} (site 1b). The values of \textit{z} and \textit{z'} for the different models are listed in table AI.}
\begin{center}
\begin{tabular}{lllllllllllllllllllllllll}
\hline\hline
\ \\
\textit{\textbf{P4/mmm}}  & & & & & & & & & & & & & &\textit{\textbf{P4mm}}          \ \\
\ \\
\hline
\ \\
$Fe_{1}$   & (1/2,1/2,z)   & & & & & & & & & & & & &  $Fe_{1}$        &(1/2,1/2,z)     \ \\
$Fe_{2}$   & (1/2,1/2,-z)  & & & & & & & & & & & & &                  &                 \ \\
$Cu_{1}$   & (1/2,1/2,z')  & & & & & & & & & & & & &  $Cu_{1}$        &(1/2,1/2,z')     \ \\
$Cu_{2}$   & (1/2,1/2,-z') & & & & & & & & & & & & &                  &                 \ \\
\ \\
\hline\hline
\end{tabular}
\end{center}
\end{table}

\begin{table}[ht!]
\caption{\label{tab:tab-generators}Labels of the generators of the space groups \textit{P4/mmm} and \textit{P4mm}, their symbols in Hermann-Maugin notation and their action on general positions within the unit cell.}
\begin{center}
\begin{tabular}{lllllllllllllllll}
\hline\hline
\ \\
\textit{\textbf{P4/mmm}}  & & & & & &\textit{\textbf{P4mm}}        \ \\
\ \\
\hline
\ \\
\textbf{2$_z$} = 2 (0,0,z); (-x, -y, z)   & & & & & &\textbf{2$_z$} = 2 (0,0,z); (-x, -y, z)     \ \\
\textbf{2$_y$} = 2 (0,y,z); (-x, y, -z)   & & & & & &\textbf{4$^+$} = 4$^+$ (0,0,z); (-y, x, z)  \ \\
\textbf{4$^+$} = 4$^+$ (0,0,z); (-y, x, z)& & & & & &\textbf{m$_{xz}$} = m (x,0,z); (x, -y, z)   \ \\
\textbf{i}     = -1 (0,0,0); (-x, -y, -z) & & & & & &    \ \\
\ \\
\hline\hline
\end{tabular}
\end{center}
\end{table}

\subsection{Commensurate magnetic structure}

For the commensurate magnetic propagation vector \textbf{k$_c$} = (1/2, 1/2, 1/2), \textbf{k$_c$} is equivalent to -\textbf{k$_c$} for both $P4/mmm$ and $P4mm$. Moreover, the little group G$_\textbf{k}$ coincides with G for the two space groups. The generators of G$_\textbf{k}$ are listed in Tab.~\ref{tab:tab-generators} and the characters of the \textit{irreps}, as calculated by the program BasiReps \cite{FullProf_93}, are displayed in Tab.~\ref{tab:tab-charcomm}.

\begin{table}[ht!]
\caption{\label{tab:tab-charcomm}Characters of the irreducible representations for the space groups \textit{P4/mmm} and \textit{P4mm} and the commensurate propagation vector \textbf{k$_c$} = (1/2, 1/2, 1/2). For both space groups G$_\textbf{k}$ coincides with G. Only the characters corresponding to the generators of G$_\textbf{k}$ are shown.}
\begin{center}
\begin{tabular}{lrrrrccccccccccccclrrr}
\hline\hline
\ \\
\textit{\textbf{P4/mmm}}  & & & & & & & & & & & & & & & & & &\textit{\textbf{P4mm}} &  &  &  \ \\
\ \\
\hline
\ \\
 &\textbf{2$_z$} &\textbf{2$_y$} &\textbf{4$^+$} &\textbf{i} & & & & & & & & & & & & &  & &\textbf{2$_z$} &\textbf{4$^+$} &\textbf{m$_{xz}$}\ \\
\ \\
\hline
\ \\
$\Gamma_{1}^{c}$    & 1    & 1    &  1    & 1    & & & & & & & & & & & & &   & $\Gamma_{1}^{c}$    & 1    & 1    &  1   \ \\
$\Gamma_{2}^{c}$    & 1    & 1    &  1    & -1   & & & & & & & & & & & & &   & $\Gamma_{2}^{c}$    & 1    & 1    &  -1  \ \\
$\Gamma_{3}^{c}$    & 1    & 1    &  -1   & 1    & & & & & & & & & & & & &   & $\Gamma_{3}^{c}$    & 1    & -1   &  1   \ \\
$\Gamma_{4}^{c}$    & 1    & 1    &  -1   & -1   & & & & & & & & & & & & &   & $\Gamma_{4}^{c}$    & 1    & -1   &  -1  \ \\
$\Gamma_{5}^{c}$    & 1    & -1   &  1    & 1    & & & & & & & & & & & & &   & $\Gamma_{5}^{c}$    & -2   & 0    &  0   \ \\
$\Gamma_{6}^{c}$    & 1    & -1   &  1    & -1   & & & & & & & & & & & & &   &               &      &      &      \ \\
$\Gamma_{7}^{c}$    & 1    & -1   &  -1   & 1    & & & & & & & & & & & & &   &               &      &      &      \ \\
$\Gamma_{8}^{c}$    & 1    & -1   &  -1   & -1   & & & & & & & & & & & & &   &               &      &      &      \ \\
$\Gamma_{9}^{c}$    & -2   & 0    &  0    & -2   & & & & & & & & & & & & &   &               &      &      &      \ \\
$\Gamma_{10}^{c}$   & -2   & 0    &  0    & 2    & & & & & & & & & & & & &   &               &      &      &      \ \\
\ \\
 \hline\hline
\end{tabular}
\end{center}
\end{table}

\begin{table}[ht!]
\caption{\label{tab:tab-BFcentrocomm}Basis functions for the Fe atoms at the site (2h) of the space group \textit{P4/mmm} for the commensurate propagation vector \textbf{k$_c$} = (1/2, 1/2, 1/2). They also apply to the Cu atoms, which occupy the same crystallographic site with a different z coordinate.}
\begin{center}
\begin{tabular}{lcccccccccccc}
\hline\hline
\ \\
\textit{\textbf{P4/mmm}} (\textbf{k$_c$}) &     &                        &                         \ \\
\ \\
\hline
\ \\
IR               &Basis         & & & & $Fe_{1}$                 & & & & $Fe_{2}$       \ \\
                 &vectors       & & & & (1/2, 1/2, z)            & & & & (1/2, 1/2, -z)  \ \\
\ \\
\ \\
\hline
\ \\
$\Gamma_{1}^{c}$  & -           & & & & -                       & & & &  -        \ \\
\ \\
$\Gamma_{2}^{c}$  &$V_{1}^{1}$  & & & &Re: [0,0,1]               & & & & [0,0,-1]   \ \\
\ \\
$\Gamma_{3}^{c}$  & -           & & & & -                       & & & &  -        \ \\
\ \\
$\Gamma_{4}^{c}$  & -           & & & & -                       & & & &  -         \ \\
\ \\
$\Gamma_{5}^{c}$  & -           & & & & -                       & & & &  -         \ \\
\ \\
$\Gamma_{6}^{c}$  & -           & & & & -                       & & & &  -         \ \\
\ \\
$\Gamma_{7}^{c}$  &$V_{1}^{7}$  & & & &Re: [0,0,1]              & & & & [0,0,1]      \ \\
\ \\
$\Gamma_{8}^{c}$  & -           & & & & -                       & & & &  -          \ \\
\ \\
$\Gamma_{9}^{c}$  &$V_{1}^{9}$  & & & &Re: [1,0,0]              & & & & [-1,0,0]     \ \\
                  &$V_{2}^{9}$  & & & &Re: [0,-1,0]             & & & & [0,1,0]      \ \\
\ \\
$\Gamma_{10}^{c}$ &$V_{1}^{10}$ & & & &Re: [0,1,0]              & & & & [0,1,0]      \ \\
                  &$V_{2}^{10}$ & & & &Re: [1,0,0]              & & & & [1,0,0]       \ \\

\ \\
 \hline\hline
\end{tabular}
\end{center}
\end{table}

\begin{table}[ht!]
\caption{\label{tab:tab-charincomm}Characters of the irreducible representations for the space groups \textit{P4/mmm} and \textit{P4mm} and the ICM propagation vector \textbf{k$_i$} = (1/2, 1/2, \textit{k$_z$}). For both space groups G$_\textbf{k}$ = \textit{P4mm}. Only the characters corresponding to the generators of G$_\textbf{k}$ are shown.}
\begin{center}
\begin{tabular}{lrrrccccccccccccclrrr}
\hline\hline
\ \\
\textit{\textbf{P4/mmm}}             & & & &   & & & & & & & & & & & &  &\textit{\textbf{P4mm}} &  &  &  \ \\
\ \\
\hline
\ \\
 &\textbf{2$_z$} &\textbf{4$^+$} &\textbf{m$_{xz}$} & & & & & & & & & & & & & & &\textbf{2$_z$} &\textbf{4$^+$} &\textbf{m$_{xz}$}\ \\
\ \\
\hline
\ \\
$\Gamma_{1}^{i}$    & 1    & 1    &  1       & & & & & & & & & & & & &   & $\Gamma_{1}^{i}$    & 1    & 1    &  1   \ \\
$\Gamma_{2}^{i}$    & 1    & 1    &  -1      & & & & & & & & & & & & &   & $\Gamma_{2}^{i}$    & 1    & 1    &  -1  \ \\
$\Gamma_{3}^{i}$    & 1    & -1   &  1       & & & & & & & & & & & & &   & $\Gamma_{3}^{i}$    & 1    & -1   &  1   \ \\
$\Gamma_{4}^{i}$    & 1    & -1   &  -1      & & & & & & & & & & & & &   & $\Gamma_{4}^{i}$    & 1    & -1   &  -1  \ \\
$\Gamma_{5}^{i}$    & -2   & 0    &  0       & & & & & & & & & & & & &   & $\Gamma_{5}^{i}$    & -2   & 0    &  0   \ \\

\ \\
 \hline\hline
\end{tabular}
\end{center}
\end{table}

\begin{table}[ht!]
\caption{\label{tab:tab-BFrest}Basis functions for Fe atoms at the site (2h) of the space group \textit{P4/mmm} for the ICM propagation vector \textbf{k$_i$} = (1/2, 1/2, \textit{k$_z$}). The table applies to each one of the two orbits in which the site (2h) splits for \textbf{k$_i$}. The same table is obtained for the site (1b) of the space group \textit{P4mm} with \textit{both} \textbf{k$_c$} = (1/2, 1/2, 1/2) and \textbf{k$_i$} = (1/2, 1/2, \textit{k$_z$}) }
\begin{center}
\begin{tabular}{lcccccccccccccccccccc}
\hline\hline
\ \\
\textit{\textbf{P4/mmm}} (\textbf{k$_i$}) &      &                   \ \\
\textit{\textbf{P4mm}} (\textbf{k$_c$} and \textbf{k$_i$}) &      &                   \ \\
\ \\
\hline
\ \\
IR                   & & & & Basis         & & & &$Fe_{1}$             \ \\
                     & & & & vectors       & & & &(1/2, 1/2, z)  \ \\
\ \\
\hline
\ \\
$\Gamma_{1}^{c}$ and $\Gamma_{1}^{i}$    & & & &-                                       \ \\
\ \\
$\Gamma_{2}^{c}$ and $\Gamma_{2}^{i}$    & & & &$V_{1}^{2}$   & & & &Re: [0,0,1]    \ \\
\ \\
$\Gamma_{3}^{c}$ and $\Gamma_{3}^{i}$    & & & &-                                       \ \\
\ \\
$\Gamma_{4}^{c}$ and $\Gamma_{4}^{i}$    & & & &-                                       \ \\
\ \\
$\Gamma_{5}^{c}$ and $\Gamma_{5}^{i}$    & & & &$V_{1}^{5}$   & & & &Re: [1,0,0]    \ \\
                     & & & &              & & & &Im: [0,-1,0]                 \ \\
\ \\
                     & & & &$V_{2}^{5}$   & & & &Re: [-1,0,0]   \ \\
                     & & & &              & & & &Im: [0,-1,0]               \ \\
\ \\
\hline\hline
\end{tabular}
\end{center}
\end{table}

A reducible representation $\Gamma$ can be constructed using the transformation properties of the magnetic moment components at the Fe sites. The decomposition of $\Gamma$ in terms of \textit{irreps} for G$_\textbf{k}$ = $P4/mmm$ is:

\begin{equation}
\Gamma = \Gamma_{2}^{c} + \Gamma_{7}^{c} + \Gamma_{9}^{c} + \Gamma_{10}^{c}
\label{Eq_Heisenberg}
\end{equation}

and for G$_\textbf{k}$ = $P4mm$:

\begin{equation}
 \Gamma = \Gamma_{2}^{c} + \Gamma_{5}^{c}
\label{Eq_Heisenberg}
\end{equation}

The basis vectors obtained using the projector operator method for each one of the \textit{irreps} appearing in the decomposition are listed in Tables~\ref{tab:tab-BFcentrocomm} and ~\ref{tab:tab-BFrest}. For $P4/mmm$ the Fe magnetic moments can be along the $\textbf{c}$ axis ($\Gamma_{2}^{c}$ and $\Gamma_{7}^{c}$) or within the $\textbf{ab}$ plane ($\Gamma_{9}^{c}$ and $\Gamma_{10}^{c}$). For $\Gamma_{2}^{c}$ and $\Gamma_{9}^{c}$ the coupling between the moments at the Fe$_1$ and Fe$_2$ sites is antiferromagnetic (AFM) whereas for $\Gamma_{7}^{c}$ and  $\Gamma_{10}^{c}$ it is ferromagnetic (FM).

\begin{figure*}[tbh]
\includegraphics[keepaspectratio=true,width=135 mm]{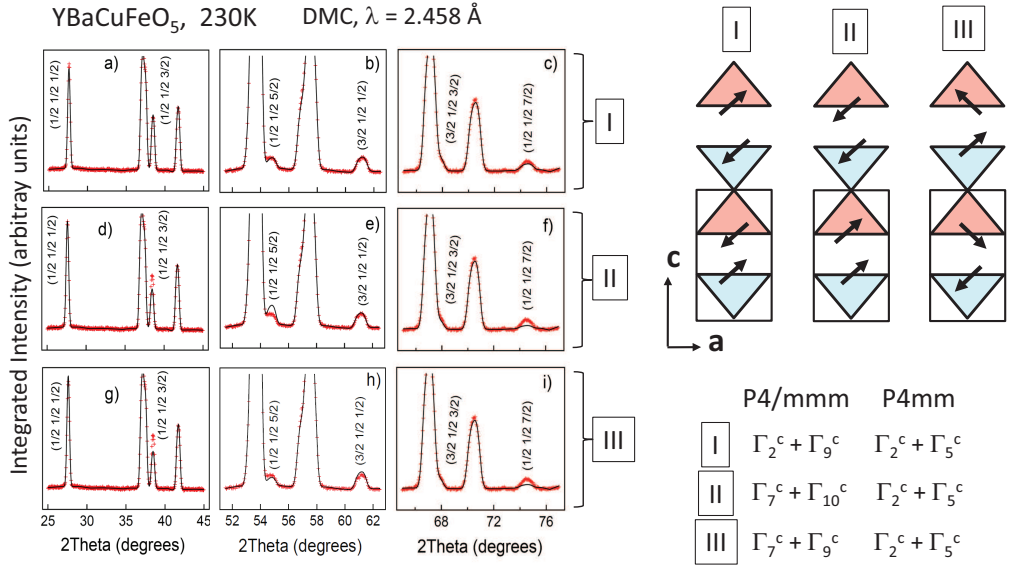}
\vspace{-3mm}
\caption{(Color online) Left panel: Selected regions of the RT neutron powder diffraction pattern of YBaCuFeO$_5$ measured at DMC with $\lambda$  = 2.458${\AA}$. Red crosses indicate experimental data, and continuous lines the fits carried out using the models for the magnetic structure schematized in the right panel. The \textit{irreps} describing the different magnetic orders in the sapce groups \textit{P4/mmm} and \textit{P4mmm} are also indicated.}
\label{fig:DMC_fits_collinear}\vspace{-5mm}
\end{figure*}

In the case of non-centrosymmetric $P4mm$ the Fe moments can only be along the $\textbf{c}$ axis for $\Gamma_{2}^{c}$. For $\Gamma_{5}^{c}$, which is two-dimensional and complex, the orientation of the Fe moment can't be obtained in a direct way. However, by imposing the condition that the magnetic moments should be real, it is easy to show that any linear combination of basis vectors gives rise to magnetic moments lying within the $\textbf{ab}$ plane. To be noted is that for both \textit{irreps}, the phase between the moments at sites 1 and 2 can not be derived from symmetry arguments and has to be obtained from the fit of the experimental data.

Although symmetry arguments require the Cu/Fe magnetic moments to be either along the \textbf{c} axis or within the \textbf{ab} plane, no satisfactory fits could be achieved with these restrictions. Combining two irreps, two collinear arrangements and a commensurate helix with pitch = $\frac{\pi}{2}$ give reasonably good agreements. They are shown in Fig.~\ref{fig:DMC_fits_collinear} together with the associated \textit{irreps} for each space group. As mentioned in section V, the three models are indistinguishable if Cu and Fe are sitting at z=1/4 and 3/4. Using the actual atomic coordinates it is however possible to distinguish them. This is shown in Fig.~\ref{fig:DMC_fits_collinear}, where the results obtained for the three models are displayed. For the Fe/Cu distribution we used the Model 4 of Table~\ref{tab:tab-structure}, although identical conclusions were obtained for the other disordered arrangement (Model 2) and sligtly different values of the Fe and Cu magnetic moments.

\begin{figure*}[tbh]
\includegraphics[keepaspectratio=true,width=155mm]{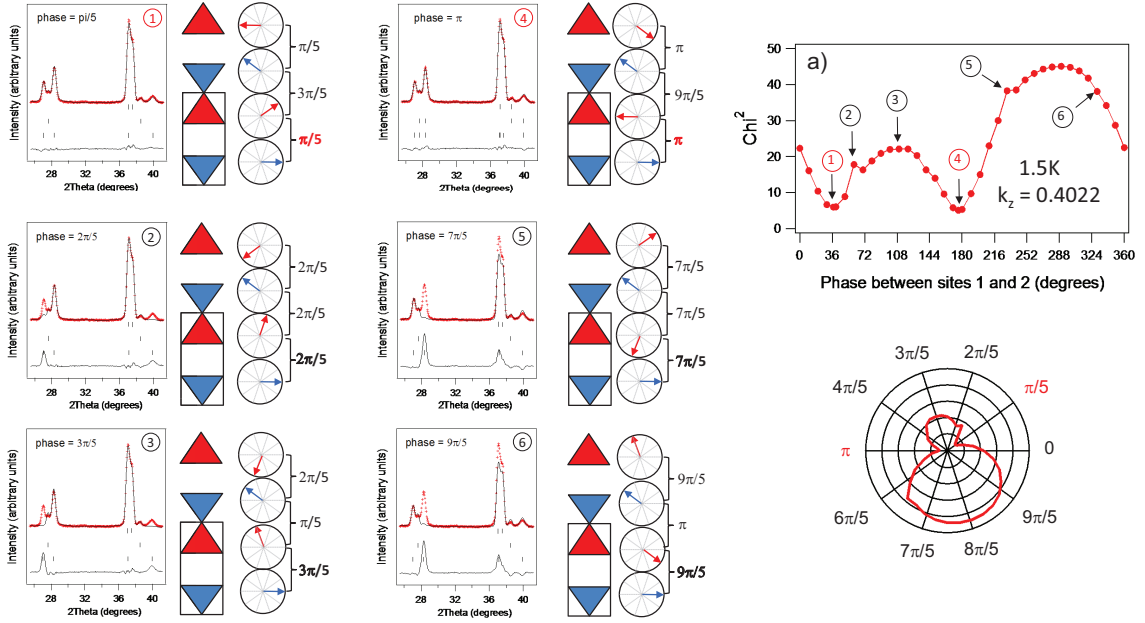}
\vspace{-3mm}
\caption{(Color online) Upper pannels: variation of Chi2 as a function of the phase between the magnetic moments at the 1 (blue) and 2 (red) sites. Lower panels: fits of the 1.5K data recorded on DMC with $\lambda$  = 2.45${\AA}$ using some representative values of the phases. Red crosses: experimental data. Black continuous lines: calculated patterns. The three rows of ticks indicate the positions of Bragg reflections of (up to down) the nuclear, the AFM commensurate and the AFM ICM phases. Particularly unstable fits ("spikes" in a) marked as (2) and (5)) are obtained for values of the phase giving rise to spirals with uniform pitch (2$\pi$/5 and 7$\pi$/5).}
\label{Fig:DMC_fits_spiral}\vspace{-5mm}
\end{figure*}

\subsection{Incommensurate magnetic structure}

For the ICM magnetic structure $\mathbf{k}_i$ = (1/2, 1/2, $k_z$) is not equivalent to $-\mathbf{k}_i$, neither for $P4/mmm$ nor for $P4mm$. In the first case the possible magnetic structures are described by two arms of the star ($\mathbf{k}_i$ and $-\mathbf{k}_i$). For $P4mm$, whose star has only one arm ($\mathbf{k}_i$), we need to add the propagation vector $\mathbf{k} = -\mathbf{k}_i$ for the description of magnetic structures with real magnetic moments. We note also that the little group G$_\textbf{k}$ = $P4mm$ is the same for both $P4/mmm$ and $P4mm$.

The characters of the \textit{irreps} of G$_\textbf{k}$ calculated by the program BasiReps \cite{FullProf_93} are displayed in Tab.~\ref{tab:tab-charincomm}. The decomposition of the reducible representation $\Gamma$ in terms of \textit{irreps} for G$_\textbf{k}$ = $P4mm$ is:

\begin{equation}
 \Gamma = \Gamma_{2}^{i} + \Gamma_{5}^{i}.
\label{Eq_Heisenberg}
\end{equation}

The basis vectors associated with the \textit{irreps} appearing in the decomposition are listed in Tab.~\ref{tab:tab-BFrest}. As in the case of the commensurate magnetic structure, the Fe$_1$ moment can only be along the $\textbf{c}$ axis for $\Gamma_{2}^{i}$ or in the \textbf{ab} plane for $\Gamma_{5}^{i}$. The phase between the moments of the ions sitting at the upper and lower pyramids can not be derived from symmetry arguments and has to be obtained from the fit of the experimental data.

Fig.~\ref{Fig:DMC_fits_spiral} shows the evolution of the goodness of the fit as a function of the phase $\phi$ between the magnetic moments at the 1 (blue) and 2 (red) sites (note that the phase between two blue (two red) sites along \textbf{c} is given by the propagation vector). Only two values of $\phi$ can reproduce the neutron powder diffraction data: $\phi$ = $\pi$ and $\phi$ = $\pi$/5. The first one involves the preservation of the AFM coupling between the red and blue sites without connecting oxygen observed in the commensurate phase and the loss of the FM coupling within the bipyramids.

In the second, the moments within the bipyramids form an angle of $\pi$/2 as in the commensurate helix of Fig.~\ref{fig:DMC_fits_collinear} whereas the perpendicular arrangement is lost between the metallic sites without linking oxygen. In view of the smooth evolution between the commensurate and ICM structures, the first model ($\phi$ = $\pi$) looks more plausible. Note also that it implies the existence of a less robust magnetic coupling within the bipyramidal units.

We tried models based on variable-moment sinusoids and constant-moment helices, either longitudinal, transverse or inclined with respect to the direction of ICM modulation vector. As for the commensurate magnetic structure, those with moments either along \textbf{c} or within the \textbf{ab} plane gave very poor agreements. The best results were obtained for inclined circular helices, although equally good refinements could be obtained with a variable moment sinusoid and values of the Fe and Cu moments slightly larger. In view of the observation of polarization only by entering the ICM phase, the helical solutions were retained.

\subsection{Calculation of exchange couplings}
\label{App_ExcCalc}
Exchange coupling constants were extracted from {\it ab initio} calculation  using  collinear spin states (with spin-orbit coupling switched off).
We applied the procedure described below \cite{Wanghbo2011} assuming that all magnetic contributions to the energy arise from the Heisenberg hamiltonian (Eq.~$1$ in the text).

To extract the exchange coupling, $J_{AB}$, between ion $A$ and ion $B$ in the supercell, we calculate the energy for the following states:
1) all the spins in the unit cell are in the collinear commensurate spin order (i.e. the spin of ion A, $\mathbf{S}_{A0}$, is  either parallel or antiparallel to the of ion B, $\mathbf{S}_{B0}$, according to the commensurate spin order),
2) spin of ion $A$ is flipped from its direction in the previous structure,
3) spin of ion $B$ is flipped from its direction in configuration 1,
4) spin of ions $A$ and $B$ are both flipped.
Assuming the magnetic ordering to be in the $z$ direction (we note that as spin-orbit coupling is not considered all directions are equivalent) and according to Eq.~$1$, the energies per supercell of these states are
\begin{eqnarray}
\label{eq_system}
E_1\! &=&\! n J_{AB} S^z_{A0} S^z_{B0} + h_A S^z_{A0}+ h_B S^z_{B0} + \epsilon   \nonumber \\
E_2 \! &=& \! - n J_{AB} S^z_{A0} S^z_{B0} - h_A S^z_{A0} + h_B   S^z_{B0} + \epsilon \nonumber \\
E_3 \! &=& \! -n J_{AB} S^z_{A0} S^z_{B0} + h_A S^z_{A0}  - h_B S^z_{B0} + \epsilon \nonumber \\
E_4 \!&=&\! n J_{AB} S^z_{A0} S^z_{B0} - h_A S^z_{A0}  - h_B S^z_{B0}+ \epsilon, \nonumber \\
\end{eqnarray}
where,  $h_i$ is the effective field generated by all the spins in magnetic sublattices different from $A$ or $B$ and $\epsilon$ is the sum of all the contributions (magnetic and non-magnetic) not involving spins at sublattices $A$ and $B$. Moreover, $n$ is the multiplicity of equivalent bonds connecting ion A to ion B when periodic boundary conditions are taken into account  (e.g., using the four formula unit u.c. and the notation in Tab.~I: $n(J_{1,2}) =4$, $n(J_{1,3}) =1$ and  $n(J_{3,5}) =1$).
Once the energies $E_i$ are calculated {\it ab-initio}, Eqs.~\ref{eq_system} become a system of equations in the unknown values $J_{AB}$, $h_A$, $h_B$ and $\epsilon$ which is solved to extract $J_{AB}$.

Additionally, we note that, once further-neighbor couplings are considered and the unit cell used is not large enough, due to periodic boundary conditions, ions A and B might be connected by two inequivalent bonds. This happens, for example, in configuration a)  when next-nearest-neighbor couplings along c are considered. In such a case, the NNN coupling J$_{1,5}$ is not equivalent to NNN coupling J$_{1+c,5}$ as the first exchange goes through a Cu-Cu-Fe pathway while the second goes through Cu-Fe-Fe pathway.
It is not possible to extract these exchange couplings using the four formula units u.c. in Fig.~\ref{fig:Structures}a as, for such supercell, the above method would yield only the average value of the two inequivalent couplings.
Therefore, to extract separately these exchanges, it is necessary to double the supercell along the c direction.
Indeed, to separate inequivalent NNN couplings for configuration a, c and d, a supercell doubled along the c direction together with a $\Gamma$-centered k-point grid of size $8\times 8 \times 2$ was used in the calculations.

\section{Acknowledgements}

We thank the Swiss National Science Foundation (grants 200021-141334/1 and 200021-141357), and the Swiss National Center of Competence in Research MARVEL (Computational Design and Discovery of Novel Materials) for financial support. The allocation of neutron beamtime at the SINQ and computer time at the Swiss National Supercomputing Center (CSCS) are gratefully acknowledged.

\bibliographystyle{apsrev}
\vspace{-7mm}
\bibliography{YBCF_PRB}

\begin{thebibliography}{31}
\bibitem[{SM()}]{SM}
\bibinfo{note}{{J$_{\parallel \mbox{O}}$ for a Cu-Cu pair may be both FM and
  AFM but in both cases it is negligible small.}}
\end{thebibliography}

\end{document}